\documentclass[final]{svjour2}
\usepackage{graphicx}
\usepackage{rotating}
\usepackage{amssymb}
\usepackage{mathptmx}
\usepackage[numbers]{natbib}
\makeatletter
\journalname{Journal of Low Temperature Physics}

\bibpunct{}{}{,}{s}{}{,}

\begin{document}

\newcommand{\hdblarrow}{H\makebox[0.9ex][l]{$\downdownarrows$}-}
\title{Alpha-like clustering in $^{20}$Ne from a quartetting wave function approach}

%
\author{G. R\"{o}pke \and P. Schuck\and C. Xu\and Z. Ren \and M. Lyu \and B. Zhou \and Y. Funaki \and
H. Horiuchi \and A. Tohsaki \and T. Yamada}
\institute{Institut f\"{u}r Physik, Universit\"{a}t Rostock, D-18051 Rostock, Germany
\and Institut de Physique Nucl\'{e}aire, Universit\'e Paris-Sud, IN2P3-CNRS, UMR 8608, F-91406, Orsay, France \and
 Department of Physics, Nanjing University, Nanjing 210093, China \and Research Center for
   Nuclear Physics (RCNP), Osaka University, Osaka 567-0047, Japan \and Laboratory of Physics, Kanto Gakuin University, Yokohama 236-8501, Japan}

\maketitle

\begin{abstract}
Quartetting ($\alpha$-like clustering) occurs in low density matter ($\le 0.03$ fm$^{-3}$) which exists, e.g., at the surface of nuclei.
It is of interest for the $\alpha$ preformation to calculate the $\alpha$ decay of heavy nuclei such as $^{212}$Po, but also in light nuclei (e.g., $^{20}$Ne) which shows strong signatures of quartetting. 
We analyze the intrinsic structure of the $\alpha$-like
cluster and the center of mass motion of the quartet, in particular the role of Pauli blocking. 
The Thomas-Fermi model for the (daughter) core nucleus is improved introducing quasiparticle  
 nucleon states. Calculations performed for harmonic oscillator basis states show
that the effective potential for the quartet center of mass motion remains nearly constant within the core nucleus. 
The relation to the THSR (Tohsaki-Horiuchi-Schuck-R\"opke) approach is discussed.


\end{abstract}

\section{Introduction}

Nuclear systems are strongly interacting so that correlations are relevant. At high densities where the nucleons are
degenerate, a quasiparticle approach, such as shell model calculations, 
is successful to describe the properties of nuclei.
Near the saturation density ($n_{B}^{\rm sat}\approx 0.15$ fm$^{-3}$)
 nuclear matter is well described by the Fermi liquid model of Landau and Migdal,
 or the Walecka relativistic mean-field (RMF) approach. For the relation between both
 quasiparticle approaches see \cite{Matsui}. However, if the nucleon density becomes low, correlations and cluster formation
will occur. Whereas pairing occurs also in dense, degenerated matter, four-particle correlations 
and $\alpha$-like clustering appear at low baryon densities $n_B < 0.03$ fm$^{-3}$. The reasons for the disappearance  of
light clusters at increasing baryon density are self-energy shifts and Pauli blocking by the surrounding medium \cite{RMS,R}.
At low density, $\alpha$-like correlations
become relevant because of the relatively high binding energy. In contrast to the deuteron where 
the binding energy per nucleon is about 1.1 MeV, a value 7.1 MeV is observed for $^4$He, the $\alpha$ particle.

A consistent description of quartetting ($\alpha$-like correlations) has been worked out recently 
within the Tohsaki-Horiuchi-Schuck-R\"opke (THSR) approach \cite{THSR}.  This approach gives an excellent 
description of low-density 4$n$ nuclei such as $^8$Be, the Hoyle state of $^{12}$C, and excited states 
of $^{16}$O, but was applied also to more complex nuclei such as $^{20}$Ne \cite{Bo3,Bo,Bo2} as well as
4$n$ nuclei with additional nucleons \cite{Lyu}.

Another site where quartetting is important, is the preformation of  $\alpha$ particles 
at the surface of heavy nuclei.
As an example, $^{212}$Po has been considered recently within  a quartetting wave function approach
\cite{Po,Xu} which describes a quartet on top of the  $^{202}$Pb core nucleus. The formation of a pocket 
of the effective potential of the  $\alpha$ particle near the surface of the core nucleus was obtained. The preformation
factor and the half life of the decay was calculated. 
This quartetting wave function approach is presently successfully applied to 
a wider class of $\alpha$ emitting nuclei, in particular all isotopes of Po \cite{Xu1}. Excellent results for the
preformation factor are obtained. This gives some confidence in the approximations performed when 
working out a microscopic approach to the preformation factor. However, 
the approximations performed in that papers have to be investigated in detail and should be improved.

In particular, a single-nucleon description (Thomas-Fermi gas model) for the core nucleus was taken, so that 
the approach was not fully consistent. 
Clustering is suppressed in the $^{208}$Pb core nucleus because it is double magic and strongly bound.
It is well-known that shell structure effects lead to clear signatures of $\alpha$ decay, 
for instance in the Po isotopes ($Z=84$) \cite{Xu1}. 
Obviously, such effects cannot be obtained from a Thomas-Fermi gas model for the core nucleus.
Nuclear shell model calculations for $\alpha$-transition probabilities of Po isotopes have been discussed
\cite{Mang}, but clustering of particles in the nuclear surface has not been taken into account, see also
\cite{Arima,Delion09,Lovas,Mirea,Delion1,DLW}.
Improving our quartetting wave function approach, we discuss some aspects of the shell model 
where the single-nucleon states are introduced as quasiparticle states.
A striking effect of the Thomas-Fermi model is that the effective potential for the quartet 
is flat inside the core nucleus We show that this behavior is approximately fulfilled 
by shell model calculations.

A more general approach is desired where correlations in the core nucleus are consistently taken into account.
As in the THSR approach, all nucleons may exhibit correlations and clustering.
Heavy nuclei with a large number of nucleons are not tractable within the THSR approach yet. 
We discuss 
here $^{20}$Ne where both approaches can be done. Comparing results for the quartetting wave function approach
with THSR calculations we find a better understanding how to describe correlations
 in nuclear systems. 
In contrast to models which consider $\alpha$-cluster at fixed configuration 
in space, see Ref. \cite{solidalpha}, in THSR calculations the $\alpha$-like 
clusters can move as described by a container model \cite{Bo3,Bo,Bo2}.

We introduce the center of mass (c.o.m.) motion of a quartet $\{ n_\uparrow, n_\downarrow, p_\uparrow, p_\downarrow \}$ 
as a new collective degree of freedom and compare the 
wave functions for both approaches.
In addition, we outline the problem how to improve the quartetting wave function approach to get a consistent description of 
cluster formation in a clustered medium.
In particular, the following approximations are essential:\\
(i) For the Pauli blocking term, a local approximation has been performed so that the intrinsic wave function changes abruptly 
from an $\alpha$-like cluster to a product of uncorrelated single-particle states as soon as the nucleon density $n_B$ 
at the c.o.m. position $R$ of the quartet exceeds a critical value  $n_B^{\rm Mott}=0.02917$ fm$^{-3}$. As a consequence, 
the effective potential $W(R )$ for the c.o.m. motion shows a kink at the critical radius $r_{\rm crit}$. It is clear that this sharp kink is a 
consequence of the approximation and will be smeared out when the intrinsic density distribution in the $\alpha$ particle
and the non-local behavior of the Pauli blocking are taken into account.\\
(ii) The Thomas-Fermi model is improved if the single-quasiparticle states inside the core nucleus (shell model) are 
taken into account. This has been already discussed in the former Refs.  \cite{Po,Xu} and will be advanced in the present work.\\
(iii) It is possible to include quartetting also for the core nucleons. This has been done within the THSR approach
and will be a topic for future work, in particular the special case of  $^{20}$Ne where both approaches can be performed.\\

\section{Density distribution of the core nucleus}
\label{sec:2}

We discuss two examples, the system $^{212}$Po = $^{208}$Pb + $\alpha$ which is an $\alpha$ emitter, 
and $^{20}$Ne = $^{16}$O + $\alpha$. In both cases, the core nuclei  $^{208}$Pb and  $^{16}$O are double magic 
so that we can assume that they have a very compact structure, and the additional quartet of nucleons 
on top of the double magic core is expected to show $\alpha$ like correlations. In the first case, $\alpha$ preformation 
in the surface region of the heavy nuclei has been discussed, see Ref. \cite{Po,Xu}, in the second case which is a $n\,\alpha$ nucleus,
THSR calculations \cite{Bo,Bo2} have shown that $\alpha$ like correlations occur.

To describe $\alpha$ like correlations in nuclei, we follow the quartetting wave function approach given in Refs. \cite{Po,Xu}. 
The four nucleons forming the quartet are moving under the influence of the core.
Within a mean-field approach considered here, we neglect all correlations between the nucleons of the core 
and the nucleons of the quartet so that the core nucleons are replaced by an averaged field acting on the quartet.
For the Coulomb interaction and the nucleon-nucleon interaction, the construction of the mean field is quite simple.
We have to know the density distribution of the core nucleons and perform a folding integral with the corresponding interactions. 
For the exchange terms between the nucleons of the quartet and the core nucleons, 
the introduction of a mean field is a delicate problem which will be discussed below 
in this work.

\subsection{Density distribution of the  $^{208}$Pb core nucleus and critical radius}

As already discussed in \cite{Xu}, for the density distribution of the lead nucleus we can use the empirical results obtained recently
\cite{Tarbert2014} which are parametrized by Fermi functions. 
The neutron density is
\begin{equation}
 n_{n,{\rm Pb}}( r)=0.093776\,{\rm fm}^{-3}/\{1+\exp[(r/{\rm fm}-6.7 )/0.55 \} 
 \end{equation}
 and the proton density
\begin{equation}
n_{p,{\rm Pb}}( r)=0.062895\,{\rm fm}^{-3}/\{1+\exp[(r/{\rm fm}-6.68 )/0.447 ]\}. 
\end{equation}
The $\alpha$ particle as a bound state can exist only for densities 
smaller than the Mott density $n_B^{\rm Mott}= 0.02917$ fm$^{-3}$.  
The Mott density $n_B^{\rm Mott}$ occurs
at the critical radius $r_{\rm crit} =7.4383$ fm so that $n_B(r_{\rm crit})=n_B^{\rm Mott}$.
This means that $\alpha$-like
clusters can exist only at distances $R > r_{\rm crit}$, for
smaller values of $r$ the intrinsic wave function is characterized
by the uncorrelated motion of the nucleons of the quartet. A figure showing the 
density distribution of $^{208}$Pb is found in \cite{Po,Xu}.

\subsection{Density distribution of the  $^{16}$O core nucleus and critical radius}

The density distribution for the  $^{16}$O core nucleus is needed to determine the mean field acting on the 
quartet under consideration. Recently, the expression 
\cite{Qu2011}
\begin{equation}
\label{Qu}
n^{\rm WS}_{B,{\rm O}}( r) = \frac{0.168 {\rm fm}^{-3}}{1+e^{(r/{\rm fm}-2.6)/0.45}}
\end{equation}
was given. The rms point radius is 2.6201 fm, the critical radius $r_{\rm crit} =3.302$ fm. 

\begin{figure}
\begin{center}
\includegraphics[%
  width=0.65\linewidth,
  keepaspectratio]{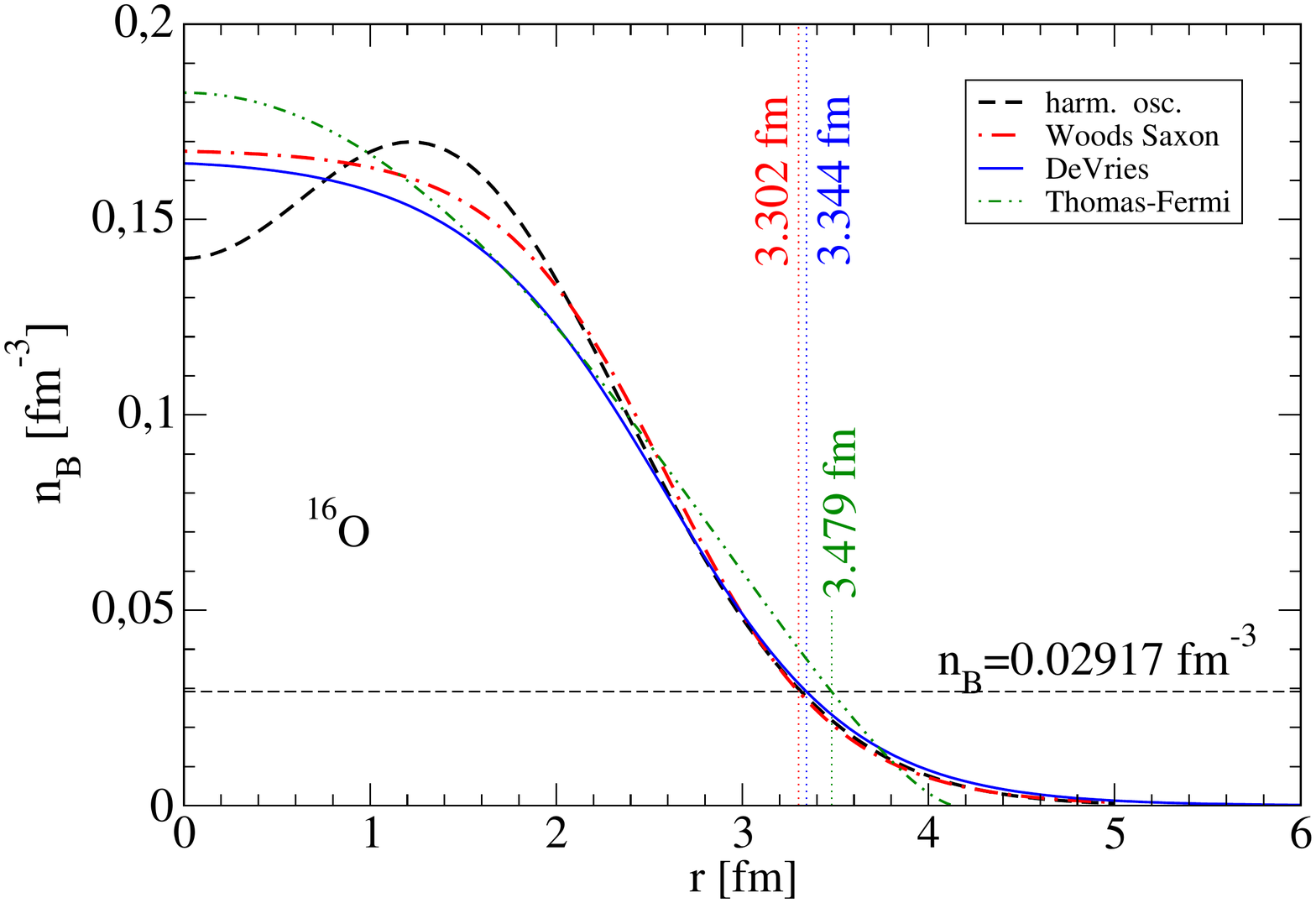}
\end{center}
\caption{Nucleon density distribution function of  the  $^{16}$O core nucleus. 
The Woods-Saxon expression (dash-dotted) \cite{Qu2011} is compared with the expression 
given by De Vries {\it et al.} (full) \cite{rms}, the Thomas-Fermi model (dash-dot-dotted),
 and the harmonic oscillator wave functions (dashed), Eq. (\ref{ho}). 
The corresponding values of $r_{\rm crit}$ for $\alpha$ formation are indicated. (Color figure online.)}
\label{fig:1}
\end{figure}

A slightly different expression according to DeVries is given in Ref. \cite{rms}.
The tails outside the nucleus (2.6 fm) nearly coincide, see Fig.~\ref{fig:1}. 
The critical radius where $\alpha$ particles disappear is $r_{\rm crit} =3.344$ fm. 
The experimental value for the rms point radius is 2.59 fm \cite{rms}.

\subsection{Harmonic oscillator model for $^{16}$O}
\label{Harmonicoscillatormodel}
Commonly used are harmonic oscillator wave functions.
Single-nucleon quasiparticle states are obtained from shell model calculations. Instead of performing such shell-model calculations,
we use here only a simple harmonic oscillator model. Instead of a mean field which is introduced self-consistently,
we use a harmonic oscillator model where the mean field is replaced by an external harmonic oscillator potential $V( r)=m \omega^2 r^2/2$.

We localize the harmonic oscillator potential at ${\bf r}=0$.
The lowest orbitals in angular momentum representation are
\begin{eqnarray}
\label{psin1}
&& \psi_{1s}( r)=\left(\frac{a}{\pi}\right)^{3/4} e^{-a r^2/2},\nonumber\\
&&  \psi_{1p}( r)=\left(\frac{a}{\pi}\right)^{3/4} e^{-a r^2/2}\left(\frac{2 a}{3}\right)^{1/2}r\,\,Y_{1m}.
\end{eqnarray}

We use a  harmonic oscillator shell model for the  $^{16}$O core nucleus.
With the parameter $a$ we have for the total nucleon density (see also \cite{Bo})
\begin{equation}
\label{ho}
n^{\rm h.o.}_{B,{\rm O}}(r )=4 \left(\frac{a}{\pi}\right)^{3/2} e^{-a r^2}+12 \left(\frac{a}{\pi}\right)^{3/2} \frac{2a}{3} r^2 e^{-a r^2}.
\end{equation}
The rms point radius $2.587$ fm is reproduced for $a=0.33619$ fm$^{-2}$.
In the tail region, this core nucleon density coincides nicely with the expressions (\ref{Qu}) given above, see also Fig. \ref{fig:1}.
The critical radius is $r_{\rm crit}^{\rm h.o.}=3.3156$ fm.

The subject of the present investigation is the comparison of the Thomas-Fermi model with the shell model.
Therefore we start from the harmonic oscillator potential $V_{^{16}{\rm O}}^{\rm h.o.}( r)= \hbar \,\omega \,a\, r^2/2
 =2.3429\, r^2$ MeV fm$^{-2}$as the single-nucleon mean field which reproduces the Gaussian orbits as well as the r.m.s radius. 
The energy levels are calculated with $\hbar \omega = 13.938$ MeV. The nucleon density follows within the Thomas-Fermi model as 
\begin{equation}
 \varrho^{\rm TF}_{^{16}{\rm O}}=\frac{2}{3 \pi^2}\left[\frac{2m}{\hbar^2}(\mu_1-2.3429\, r^2 {\rm MeV\, fm}^{-2})\right]^{3/2}.
\end{equation}
For $A=16$ we find $\mu_1=40.2048$ MeV, the critical radius is $r_{\rm crit}^{\rm TF}=3.47924$ fm.
Instead of a long-range tail, the density goes to zero at  $r=4.14249$ fm. The corresponding density profile is also shown in Fig. \ref{fig:1}. It is clear that the Thomas-Fermi model reproduces the density profile only in crude approximation. 
Oscillations and shell effects are not reproduced. In particular, the tails for large $r$ are not reproduced so that
the clustering behavior which is sensitive to the density should be treated with an adequate density profile as 
already discussed in Ref. \cite{Po}. 

\subsection{Large number harmonic oscillator model}

It is expected that with increasing number of nucleons, the general reproduction of the density distribution by the Thomas-Fermi model becomes better. 
With the large  number harmonic oscillator model we consider the question whether the shell model is approximated by the Thomas-Fermi model. 
Such a harmonic oscillator model is not realistic because it cannot describe the behavior at large distances, and because Coulomb interaction is neglected, 
but it is used here to discuss the main problem of this work, the structure of the quartet wave function in the core region $R \le r_{\rm crit}$.

As example we consider the non-interacting harmonic oscillator model, for $A=80$ ($N=Z=40$). We adopt the density at $r=0$ to the saturation value 0.15 fm$^{-3}$ and
obtain $\hbar \omega =7.15362$ MeV, $a= 0.172543$ fm$^{-2}$. The Thomas-Fermi density distribution follows as 
$n_{\rm TF}( R)=0.15 (1- 0.017491 \,r^2 /{\rm fm}^2)^{3/2}$ fm$^{-3}$ with the Fermi energy 35.2848 MeV. The critical density 0.02917 fm$^{-3}$ 
occurs at $r_{\rm crit.} = 6.16302$ fm, the density goes to zero at $r=7.56132$ fm. 

\begin{figure}
\begin{center}
\includegraphics[%
  width=0.65\linewidth,
  keepaspectratio]{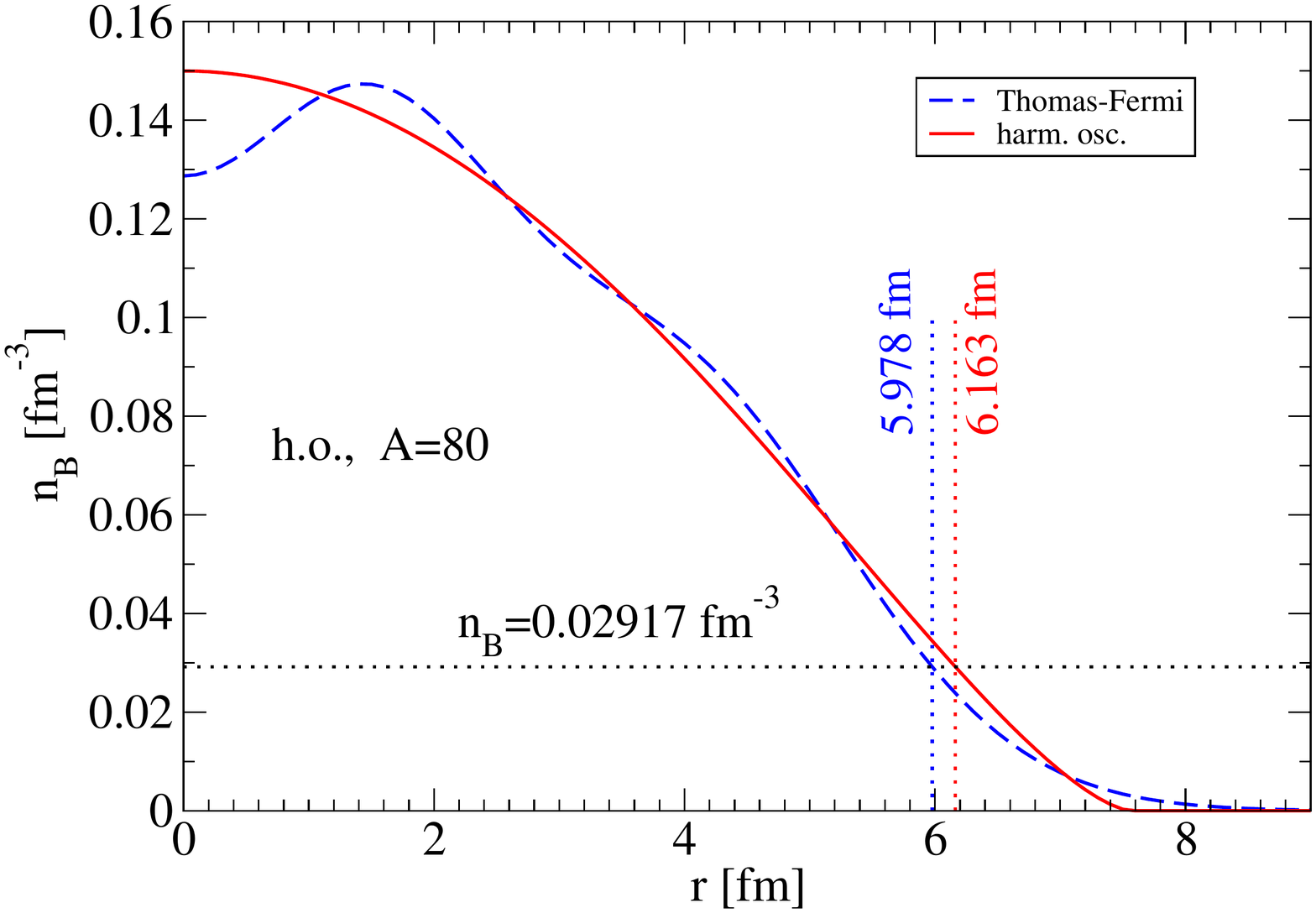}
\end{center}
\caption{Nucleon density distribution function of  non-interacting nucleons in a harmonic oscillator potential. $A=80$. Comparison of the Thomas-Fermi approximation with the exact distribution. The corresponding values of $r_{\rm crit}$ for $\alpha$ formation are indicated. (Color figure online.)}
\label{n80}    
\end{figure}

The correct density for the non-interacting  harmonic oscillator wave functions, where the states $1s,\,1p,\,1d$, $1f,\,2s,\,2p$ are occupied, is also shown in Fig. \ref{n80}. The 
 states $1d,\,1f,\,2s,\,2p$ are given in App. \ref{app:1}.
The critical value follows as $r_{\rm crit} =5.97802$ fm. The chemical potential is $5\,\, \hbar \omega=35.7681$ MeV (the middle between the highest occupied state at $9/2\,\, \hbar \omega$ and the lowest free state at $11/2\,\, \hbar \omega = 39.3449$ MeV).

The harmonic oscillator basis has the advantage that matrix elements are obtained in a simple form. 
This makes it useful for general problems and to discuss the approaches. 
We will use this advantage below in this work to compare the Thomas-Fermi model with the shell model.
In particular, we discuss the effective potential $W( R)$ for a quartet inside the core nucleus 
where Pauli blocking is described by an exchange potential.
For application to nuclei, better potentials can be used such as a combination of a Woods-Saxon potential 
with an additional term owing to the $ls$ coupling.
This may be a topic of future investigations.

\section{The quartet wave equation}

\subsection{Intrinsic and c.o.m. motion}

Within a quantum many-particle approach, the treatment of the interacting many-nucleon system needs some approximations 
that may be obtained in a consistent way from a Green functions approach.
In a first step, we can introduce the quasiparticle picture where the nucleons 
are moving independently in a mean field, described by a single-particle Hamiltonian $\hat h$,
with single-nucleon (shell) states $|n\rangle$. Because of the Pauli blocking, double occupation of the single
quasiparticle states is not allowed, what can be written using the expression 
\begin{equation}
\label{spPauli}
 \hat h=\frac{\hbar^2 p^2}{2m}+ [1 - \sum_i^{\rm occ.} |n \rangle \langle n |] V^{\rm mf}(r)
\end{equation}
so that an additional nucleon can be implemented only in the non-occupied phase space. 
The nucleon quasiparticle states are obtained from the normalized solutions of the hermitized wave equation. 

In the next 
step we go beyond the quasi-particle picture and take the full interaction within the $A_c$-particle cluster into 
account. In the case of four nucleons considered here,
we have in position space representation (see \cite{Po} and references given there)
\begin{eqnarray}
&&[E_4\!-\!\hat h_1\! -\!\hat h_2\!-\! \hat h_3\! - \!\hat h_4]\Psi({\bf r}_1 {\bf r}_2 {\bf r}_3{\bf r}_4)\!=\!\!\!
\int \!\! d^3 {\bf r}_1'\,d^3 {\bf r}_2' \langle {\bf r}_1{\bf r}_2|B \,\,V_{N-N}| {\bf r}_1'{\bf r}_2'\rangle
\Psi({\bf r}_1'{\bf r}_2'{\bf r}_3{\bf r}_4)\nonumber \\ && +
\int d^3 {\bf r}_1'\,\,d^3{\bf r}_3'  \langle {\bf r}_1{\bf r}_3|B \,\,V_{N-N}|
{\bf r}_1'{\bf r}_3'\rangle \Psi({\bf r}_1'{\bf r}_2{\bf r}_3'{\bf r}_4)
+ {\rm four\,\, further \,\,permutations.}
\label{15}
\end{eqnarray}
The six  nucleon-nucleon interaction terms contain besides the nucleon-nucleon potential $V_{N-N}$ also the 
 blocking operator $B$ that can be given in quasi-particle state representation. 
For the first term on the r.h.s. of Eq. (\ref{15}), the expression
\begin{eqnarray}
 B(1,2)=[1-f_1(\hat h_1)-f_2(\hat h_2)]  
\label{15a}
\end{eqnarray}
results neglecting any correlations. 
As in Eq. (\ref{spPauli}), the phase space occupation (we give the  internal quantum state $\nu=\sigma,\,\tau$ explicitly)
\begin{equation}
\label{occ}
f_\nu(\hat h) =\sum_n^{{\rm occ.}}| n,\nu \rangle \langle n,\nu |
\end{equation}
indicates the phase space that according to the Pauli principle is not available for an interaction process of a nucleon 
with internal quantum state $\nu$.

The mean-field $V^{\rm mf}(r)$ contains the Coulomb potential as well as the nucleon-nucleon interaction 
$V^{\rm ext}( r)$ of the core nucleus (the Hartree term is given as folding integral with the corresponding densities). The Pauli blocking terms are not easily treated 
as discussed in the following section.

A main aspect of the cluster approach is the introduction of the center-of-mass (c.o.m.) motion $\bf R$ 
as new collective degree of freedom, and ${\bf s}_j=\{\bf S,s,s'\}$ for the intrinsic motion.
We use Jacobi-Moshinsky coordinates for the quartet nucleons:
\begin{eqnarray}
\label{Jacobi}
 &&{\bf r}_{n,\uparrow}={\bf R}+{\bf S}/2+{\bf s}/2, \qquad \,\,{\bf p}_{n,\uparrow}={\bf P}/4+{\bf Q}/2+{\bf q}, \nonumber\\
&&{\bf r}_{n,\downarrow}={\bf R}+{\bf S}/2-{\bf s}/2, \qquad \,\,{\bf p}_{n,\downarrow}={\bf P}/4+{\bf Q}/2-{\bf q}, \nonumber\\
 &&{\bf r}_{p,\uparrow}={\bf R}-{\bf S}/2+{\bf s'}/2, \qquad {\bf p}_{p,\uparrow}={\bf P}/4-{\bf Q}/2+{\bf q}', \nonumber\\
&&{\bf r}_{p,\downarrow}={\bf R}-{\bf S}/2-{\bf s'}/2, \qquad {\bf p}_{p,\downarrow}={\bf P}/4-{\bf Q}/2-{\bf q}'.
\end{eqnarray}

As shown in \cite{Po}, the normalized quartet wave function in  Jacobi coordinates,
\begin{equation}
\int d^3R\,\int d^9s_j\,|\Phi({\bf R},{\bf s}_j)|^2 =1,
\end{equation}
can be decomposed in a unique way
\begin{equation}
\label{4}
\Phi({\bf R},{\bf s}_j)=\varphi^{{\rm intr}}({\bf s}_j,{\bf R})\,\psi({\bf R})
\end{equation}
(up to a phase factor) with the individual normalizations 
\begin{equation}
\label{normS}
\int d^3R\,|\psi({\bf R})|^2=1
\end{equation}
and for each ${\bf R}$
\begin{equation}
\label{normint}
\int d^9s_j |\varphi^{{\rm intr}}({\bf s}_j,{\bf R})|^2=1\,.
\end{equation}

The Hamiltonian of a cluster 
may be written as 
\begin{equation}
H=\left(-\frac{\hbar^2}{8m} \nabla_R^2+T[\nabla_{s_j}]\right)\delta^3({\bf R}-{\bf R}')\delta^3({\bf s}_j-{\bf s}'_j)
+V({\bf R},{\bf s}_j;{\bf R}',{\bf s}'_j)
\end{equation}
with the kinetic energy of the c.o.m. motion and the kinetic energy of the internal motion of the cluster, $T[\nabla_{s_j}]$. 
The interaction $V({\bf R},{\bf s}_j;{\bf R}',{\bf s}'_j)$  contains the mutual interaction $V_{ij}({\bf r}_i,{\bf r}_j,{\bf r}'_i,{\bf r}'_j)$ 
between the particles 
as well as the 
interaction with an external potential (for instance, the potential of the core nucleus). 

For the c.o.m. motion we have the wave equation
\begin{eqnarray}
\label{9}
&&-\frac{\hbar^2}{8m} \nabla_R^2\psi({\bf R})-\frac{\hbar^2}{Am}\int d^9s_j \varphi^{{\rm intr},*}({\bf s}_j,{\bf R}) 
[\nabla_R \varphi^{{\rm intr}}({\bf s}_j,{\bf R})][\nabla_R\psi({\bf R})]-
\\ &&
-\frac{\hbar^2}{8m}\int\!\! d^9s_j \varphi^{{\rm intr},*}({\bf s}_j,{\bf R}) 
[ \nabla_R^2 \varphi^{{\rm intr}}({\bf s}_j,{\bf R})] \psi({\bf R})
+\!\!\int \!\! d^3R'\,W({\bf R},{\bf R}')  \psi({\bf R}')\!=\!E\,\psi({\bf R})\,\nonumber 
\end{eqnarray}
with the c.o.m. potential
\begin{eqnarray}
\label{9c}
W({\bf R},{\bf R}')&=&\int d^9s_j\,d^9s'_j\,\varphi^{{\rm intr},*}({\bf s}_j,{\bf R}) \left[T[\nabla_{s_j}]
\delta^3({\bf R}-{\bf R}')\delta^9({\bf s}_j-{\bf s}'_j)\right.\nonumber \\&&\left.
+V({\bf R},{\bf s}_j;{\bf R}',{\bf s}'_j)\right]
\varphi^{{\rm intr}}({\bf s}'_j,{\bf R}')\,.
\end{eqnarray}
For the intrinsic motion we find the wave equation
\begin{eqnarray}
\label{10}
&&-\frac{\hbar^2}{4m}  \psi^*({\bf R}) [\nabla_R\psi({\bf R})]
[\nabla_R \varphi^{{\rm intr}}({\bf s}_j,{\bf R})]
-\frac{\hbar^2}{8m}  |\psi({\bf R})|^2
\nabla_R^2 \varphi^{{\rm intr}}({\bf s}_j,{\bf R})
\nonumber \\ &&
+\int d^3R'\,d^9s'_j\, \psi^*({\bf R}) \left[T[\nabla_{s_j}]
\delta^3({\bf R}-{\bf R}')\delta^9({\bf s}_j-{\bf s}'_j)\right.\nonumber \\&& \left.
+V({\bf R},{\bf s}_j;{\bf R}',{\bf s}'_j)\right]
 \psi({\bf R}')\varphi^{{\rm intr}}({\bf s}'_j,{\bf R}')=F({\bf R}) \varphi^{{\rm intr}}({\bf s}_j,{\bf R})\,.
\end{eqnarray}
The respective c.o.m. and intrinsic Schr\"odinger
equations are coupled by contributions containing the expression
$\nabla_R \varphi^{{\rm intr}}({\bf s}_j,{\bf R})$ which will be
neglected in the present work. This expression disappears in homogeneous matter.
No investigations of such gradient terms have performed yet for inhomogeneous systems.

\subsection{The c.o.m. potential $W({\bf R})$ and local-density Pauli blocking term}
We emphasize that we should allow for non-local interactions. In particular, the Pauli blocking considered below
is non-local. Also the nucleon-nucleon interaction can be taken as non-local potential. 
To simplify the calculations, often local approximations are used,
\begin{equation}
W({\bf R},{\bf R}')\approx W({\bf R})\delta^3({\bf R}-{\bf R}')\qquad W({\bf R})=W^{\rm ext}({\bf R})+W^{\rm intr}({\bf R}).
\end{equation}
$W^{\rm ext}({\bf R})=W^{\rm mf}({\bf R})$ is the contribution of external 
potentials, here the mean field of the core nucleons.
The interaction within the cluster according  Eq. (\ref{10}) gives the contribution $W^{\rm intr}({\bf R})$.

The intrinsic wave equation (\ref{10})  describes in the zero density limit
the formation of an $\alpha$ particle with binding energy $B_\alpha= 28.3$
MeV. For homogeneous matter, the binding energy will be reduced
because of Pauli blocking. In the zero temperature case considered
here, the shift of the binding energy is determined by the baryon
density $n_B=n_n+n_p$, i.e. the sum of the neutron density $n_n$
and the proton density $n_p$. Furthermore, Pauli blocking depends on the asymmetry given by
the proton fraction $n_p/n_B$ and the c.o.m. momentum ${\bf P}$ of the
quartet. Neglecting the weak dependence on the
asymmetry, for ${\bf P}=0$ the density dependence of the Pauli
blocking term
\begin{equation}
\label{WPauli}
 W^{\rm Pauli}(n_B)\approx 4515.9\, {\rm MeV\, fm}^3 n_B -100935\, {\rm MeV\, fm}^6 n_B^2+1202538\, {\rm MeV\, fm}^9 n_B^3
\end{equation}
was found in \cite{Po}, as a fit formula valid in the density  
region $n_B \le 0.03$ fm$^{-3}$ with relative error below 1\%.
In particular, the bound state is dissolved and merges with the continuum
of scattering states at the Mott density $n_B^{\rm Mott}= 0.02917$ fm$^{-3}$.
For the intrinsic wave function of the quartet we can assume an $\alpha$-like Gaussian
to describe the bound state. The width parameter of the free $\alpha$ particle is only 
weakly changed when approaching the Mott density, see \cite{Po}.

Below  the Mott density, $n_B \le n_B^{\rm Mott}$, the localized potential
\begin{equation}
\label{WeffR}
W({\bf R})=W^{\rm ext}({\bf R})-B_\alpha+ W^{\rm Pauli}[n_B({\bf R})]
\end{equation}
can be used as approximation. $W^{\rm ext}({\bf R})=W^{\rm mf}({\bf R})$ is the contribution of external 
potentials, here the mean field of the core nucleons. The intrinsic energy of the quartet for densities above the critical one is a minimum if all four nucleons are at the Fermi energy,
for symmetric matter $W^{\rm intr}({\bf R})=4 E_F[n_B({\bf R})]$, with the Fermi energy $E_F(n_B)=(\hbar^2/2m) (3 \pi^2n_B/2)^{2/3}$.

\subsection{The mean-field Coulomb and N - N potentials}
\label{mf}

Having the nucleon densities of the core nucleus to our disposal, the mean fields are easily calculated.
The mean-field contribution $W^{\rm mf}({\bf R})$ is obtained by double folding the density distribution of the  core 
nucleus and the intrinsic density distribution of the quartet at c.o.m. position $\bf R$ with the interaction potential. 
For the bound quartet, an  $\alpha$-like Gaussian has been taken. 

For the nucleon-nucleon contribution, 
a parametrized effective nucleon interaction 
$V_{NN}(s/{\rm fm})=c\, \exp(-4 s)/(4 s)-d\, \exp(-2.5 s)/(2.5 s)$ can be used which is motivated by the M3Y interaction \cite{M3YReview}, 
$s$ denotes the distance of nucleons. The parameters $c, d$ are adapted 
to reproduce known data.
For the lead core nucleus case, see \cite{Po,Xu,Xu1}. For the oxygen core nucleus,
parameter values $c,d$ are given below in Sec. \ref{TF20Ne}.
As also known from other mean-field approaches, we should fit the mean field to measured data.

For the Coulomb interaction we calculate
\begin{equation}
\label{VCoul}
V^{\rm Coul}_{\alpha - {\rm O}}( R) = \int d^3 r_1 \int d^3 r_2 \rho_{{\rm O}} ({\bf r}_1) \rho_\alpha ({\bf r}_2) 
\frac{e^2}{|{\bf R}-{\bf r}_1+{\bf r}_2|}
\end{equation}
with the charge density of the $\alpha$ nucleus according to 
\begin{equation}
\label{nqalpha}
 \rho_\alpha( r)=0.21144\,\,{\rm fm}^{-3} \,e^{-0.7024\,\, r^2/{\rm fm}^2} 
\end{equation}
which reproduces the measured rms point radius 1.45 fm, and the density distribution (\ref{Qu}) of $^{16}$O.
The convolution integral (\ref{VCoul}) is easily evaluated in Fourier representation and gives for the parameter values considered here
\begin{eqnarray}
&& V^{\rm Coul}_{\alpha - {\rm O}}( R)=\frac{16 \times 1.44}{R} {\rm MeV\,\, fm} \nonumber \\&& \times\left[{\rm Erf}(0.76829\,\, R/{\rm fm})-0.9097 \,\,(R/{\rm fm})\,\,e^{-0.22736\,\,R^2/{\rm fm}^2}\right]\,.
\end{eqnarray}

\section{Quartets in nuclei in Thomas-Fermi approximation}

\subsection{The Thomas-Fermi rule for the bound state energy}
\label{TFrule}

Our aim is to derive an equation of motion for the c.o.m. motion of the quartet. 
For this, we need the effective potential $W( {\bf R})$. Here, we discus the Thomas-Fermi model (local density approach).
There are two regions separated by the critical radius $r_{\rm crit}$ where the density of the core nucleus has the critical value $n_B(r_{\rm crit})= n_B^{\rm Mott}= 0.02917$ fm$^{-3}$. 
Then, $-B_\alpha+W^{\rm Pauli}(n_B)=4 E_F(n_{\rm crit})$, and the bound state merges with the continuum of scattering states.

For $R > r_{\rm crit}$, the mean-field contribution $W^{\rm ext}( R)$
is given by the double-folding Coulomb and $N-N$ potentials. The intrinsic part $W^{\rm intr}( R)$ 
contains the bound state energy -28.3 MeV of the free $\alpha$ particle which is shifted because of Pauli blocking. 
At $r_{\rm crit}$, the bound state merges with the continuum so that we have the condition (symmetric matter)
\begin{equation}
 W(r_{\rm crit})= W^{\rm ext}(r_{\rm crit})+4 E_F(n_{\rm crit}) =\mu_4,
\end{equation}
the intrinsic wave function changes from a bound state case to four uncorrelated quasiparticles on top of the Fermi sphere (the states below the Fermi energy are already occupied).

For $R < r_{\rm crit}$, in addition to the mean-field contribution $W^{\rm ext}( R)$ the Fermi energy 
$4 E_F[n( R)]$ appears. Within the Thomas-Fermi model, for a given potential $W^{\rm ext}( R)$ 
the density is determined by the condition that $W^{\rm ext}(R )+4 E_F[n_B( R)]$ remains a constant, here $\mu_4$.
We find the effective potential $W^{\rm TF}( R)$ which is continuous but has a kink at $r_{\rm crit}$.
It is an advantage of the Thomas-Fermi model that the condition $W^{\rm TF}( R)=\mu_4=$ const. holds 
for the entire region $R < r_{\rm crit}$ (where an uncorrelated product of single-particle states can be assumed), 
independent of the mean-field potential $W^{\rm ext}( R)$ and the corresponding density distribution. 
We analyze this property in the following section \ref{Modelcalculation}.

Whereas the Coulomb part to the external potential as well as the intrinsic part of the effective potential $W^{\rm TF}( R)$ are fixed, both parameters $c,d$ for the $N-N$ part of the external potential can be adjusted such 
that measured data are reproduced. For this, we have to formulate two conditions: 
i) The solution of the c.o.m. wave equation, neglecting decay, gives the energy eigenvalue $E_{\rm tunnel}$. 
This eigenvalue should coincide with the measured energy after decay as given by the $Q$ value.\\
ii) This value $E_{\rm tunnel}$ should coincide with the value $\mu_4$. 
Within the local density approach, this is the value the four nucleons must have to implement them into the core nucleus. We denote this condition $E_{\rm tunnel}=\mu_4$ as the Thomas-Fermi rule.

With both conditions, the parameter $c,d$ for the double folding $N - N$ interaction potential are found, 
and values for the preformation factor and the half life of the $\alpha$ decay have been obtained \cite{Xu}.
However, the measured half life of the $\alpha$ decay of $^{212}$Po was not well reproduced. 
The case of $^{20}$Ne will be discussed in 
Sec. \ref{TF20Ne}.

Besides other approximations performed to derive the effective wave equation for the c.o.m. motion of the 
quartet such as the omission of the gradient terms, 
it is the local density approximation which becomes questionable in finite nuclei. 
In contrast to homogeneous matter, the energy spectrum of the single nucleon quasiparticle states is not 
continuous but discrete, as known from the shell model. 
An additional nucleon is not implemented to a finite system at the energy of the highest occupied state 
as in the Thomas-Fermi model but into the next free level above the Fermi energy, 
which is separated by a certain gap value. Therefore, we have to remove the Thomas-Fermi rule given above, item ii),
allowing $E_{\rm tunnel} > \mu_4$. Then, we need a new condition to fix both parameters $c,d$, and, instead of 
the Thomas-Fermi rule, we can adjust the measured half life of the $\alpha$ decay. 
This procedure has been shown in \cite{Xu}, and a value $E_{\rm tunnel} - \mu_4=0.425$ MeV has been obtained for $^{212}$Po.

Note that we remain within the Thomas-Fermi model, only allowing for a gap when introducing additional nucleons 
to the core nucleus. A better implementation of the shell model should explain these gaps, 
but also the modifications for isotopic/isotonic series when crossing a magic number as discussed recently \cite{Xu1}.
We are not aiming to present a state of the art shell model calculation here, 
but outline the expected effects within a harmonic oscillator model calculation 
where all matrix elements can be calculated analytically, see Sec. \ref{Modelcalculation}.

In conclusion, within a Thomas-Fermi model (infinite matter), the energy of the internal motion of the quartet is given by the sum of the corresponding 
Fermi energies of the four nuclei if the density is above the critical density. At the critical density $n_B^{\rm Mott}=0.02917$ fm$^{-3}$
(symmetric matter), the sum of the four Fermi energies of the nucleons in the quartet is 47.37 MeV. Below the critical density, a bound state is formed
which lowers the energy of the quartet, see Fig. 1 of Ref. \cite{Po}.

\subsection{Parameter values and results of the Thomas-Fermi model for $^{20}$Ne}
\label{TF20Ne}

A well-known property of the Thomas-Fermi model is that the chemical potential $\mu_{\tau, \sigma}$, 
which characterizes the energy needed to add a single particle to the system, 
is not depending on position.
For a space dependent potential $V_{\tau, \sigma}( {\bf r})$, 
the local density $n_{\tau, \sigma}( {\bf r})$ is determined by the condition that the sum of the potential energy 
and the Fermi energy is constant,  $V_{\tau, \sigma}( {\bf r})+E_F[n_{\tau, \sigma}( {\bf r})]=\mu_{\tau, \sigma}$.
Consequently, for $R < r_{\rm crit}$ where the quartet is described by the product of four continuum states above the Fermi energy, 
within the Thomas-Fermi approach the effective potential $W( r)$ is a constant given by the sum $\mu_4=\sum_{\tau,\sigma} \mu_{\tau, \sigma}$ 
of the chemical potentials of the four constituents of the $\alpha$ particle, which are treated in this inner-core region as free nucleonic states. 
Because  $W( R)$ is continuous at $r_{\rm crit}$, this constant value of the in-core effective potential coincides with the outside value $W( R)$, $R> r_{\rm crit}$ at $R = r_{\rm crit}$.

In contrast to the $\alpha$ decay of $^{212}$Po where the $Q$ value can be used to estimate the chemical potential $\mu_4$ \cite{Po}, the  $^{20}$Ne is stable. However, 
we can use the additional binding when going from  $^{16}$O ($B(^{16}{\rm O})=127.66$ MeV) to  $^{20}$Ne ($B(^{20}{\rm Ne})=160.645$ MeV) 
adding the four nucleons. The difference fixes  
the position of the in-core effective potential $\mu_4 =B(^{16}{\rm O})-B(^{20}{\rm Ne})=-33.0$ MeV. 

Another condition is that  the solution of the Schr{\"o}dinger equation for the four-nucleon c.o.m. motion in the effective potential $W( R)$ gives the energy eigenvalue 
$E_{\alpha, {\rm bound}}$ at this value -33 MeV so that the $\alpha$-like cluster is at the 
Fermi energy $\mu_4 $ (see also the discussion in Ref. \cite{Xu}). 
Both conditions are used  to fix the parameters $c, d$.  The values $c=4650$ MeV and $d=1900$ MeV have been found.
\begin{figure}
\begin{center}
\includegraphics[%
  width=0.65\linewidth,
  keepaspectratio]{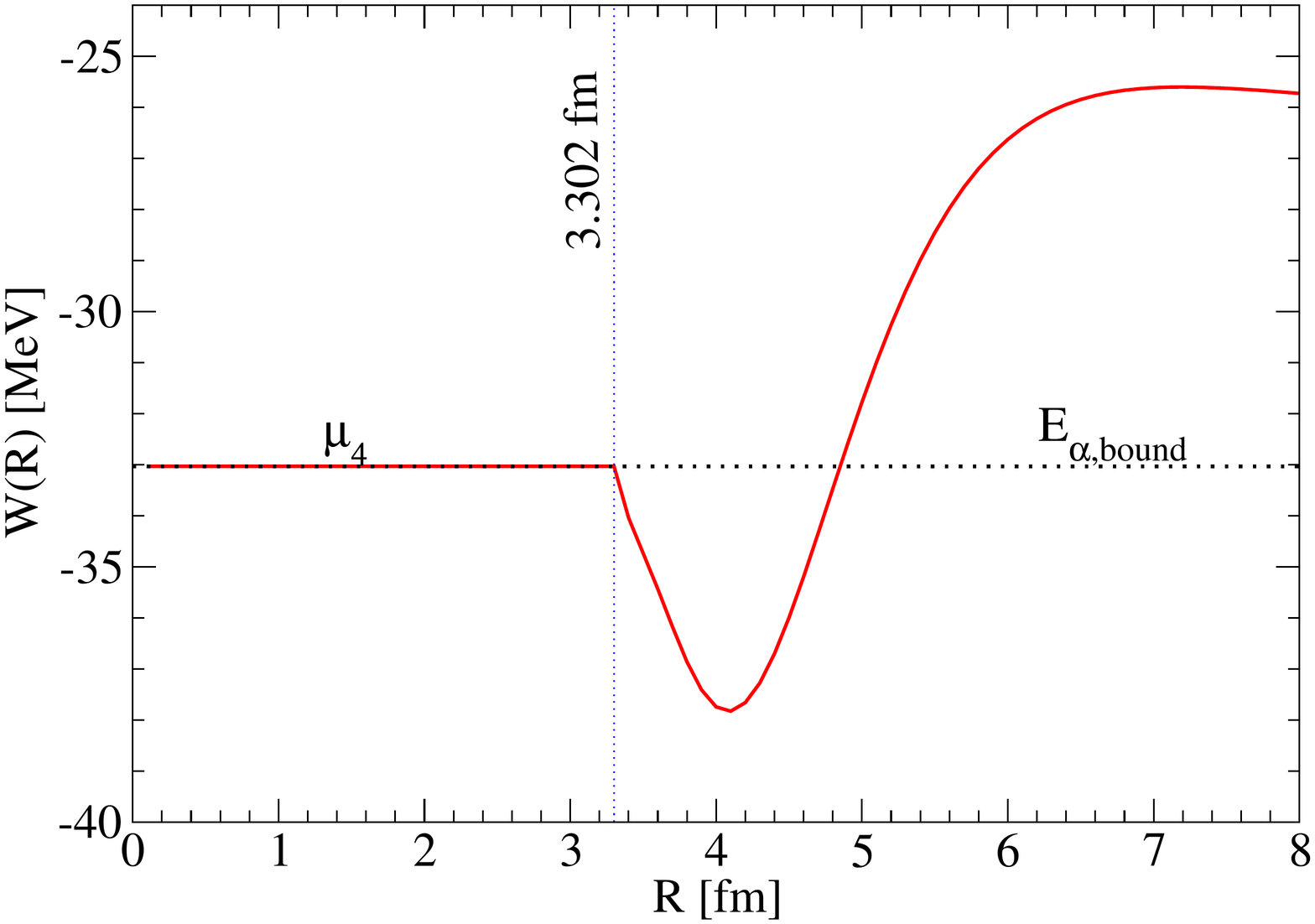}
\end{center}
\caption{Effective potential $ W( R) $ for the center of mass motion of the quartet on top of $^{16}$O. 
Thomas-Fermi model has been used. (Color figure online.)}
\label{fig:2}
\end{figure}

The resulting effective potential $ W( R) $ (\ref{WeffR}) 
for the center of mass motion of the quartet is shown in Fig. \ref{fig:2}. 
The formation of a pocket near the surface is seen which is caused by the formation of an $\alpha$-like cluster. The sharp kink at the critical radius 
$r_{\rm crit}=3.302$ fm is a consequence of the local approximation for the Pauli blocking term. A smooth behavior is expected if the finite extension of the 
$\alpha$-like cluster is taken into account so that the kink produced by the local density approximation is smeared out.

\begin{figure}
\begin{center}
\includegraphics[%
  width=0.65\linewidth,
  keepaspectratio]{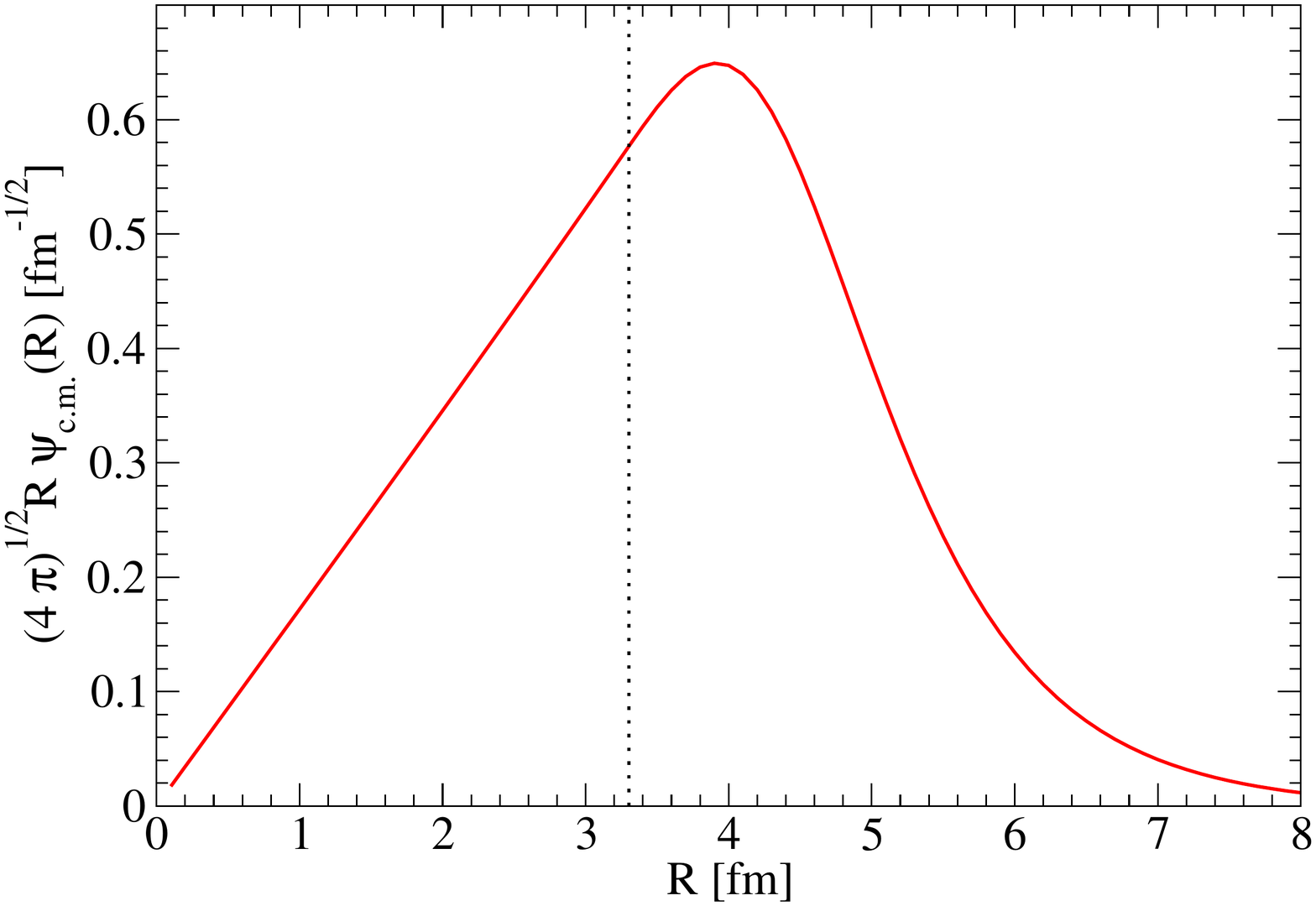}
\end{center}
\caption{Wave function for the c.o.m. motion of the quartet. A prefactor $(4 \pi)^{1/2} R$ is introduced so that the integral over $R$ of the squared quantity is normalized to 1. (Color figure online.)}
\label{fig:3}
\end{figure}

The wave function for the quartet center of mass motion $\psi_{\rm c.o.m.}( R)$ is found as solution of the Schr\"odinger equation, mass $4 m$, 
with the potential $ W( R) $. The energy eigenvalue is -33.0 MeV. A graph of $(4 \pi)^{1/2} R\, \psi_{\rm c.o.m.}( R)$ is shown in Fig. \ref{fig:3}.
The normalization is $4 \pi \int_0^\infty R^2 \psi^2_{\rm c.o.m.}( R) dR =1$. Integrating from 0 to $r_{\rm crit}=3.302$ fm, the part of the quartet 
where the internal structure is the product of free states, comes out at 0.3612. The remaining part where the internal structure is given by an $\alpha$-like
bound state is 0.6388.

As a result, we calculate the rms point radius for our solution of $^{20}$Ne as a quartet on top of the $^{16}$O core nucleus:
\begin{equation}
 {\rm rms}^2(^{20}{\rm Ne})= \frac{4 \pi}{20} \int_0^\infty r^4 \left[n^{\rm WS}_{B,{\rm O}}( r)+4 \psi^2_{\rm c.o.m.}(r)\right] dr =(2.8644\,\, {\rm fm})^2
\end{equation}
(the internal formfactor of the $\alpha$ cluster was not taken into account). The value rms$(^{20}{\rm Ne})=2.8644$ fm comes out, 
which is in good agreement with the experimental rms point radius 2.87 fm.

\section{Shell model calculation: Harmonic oscillator basis}
\label{Modelcalculation}

We are interested in a better approach which takes the discrete level structure of the core nucleus into account.
The harmonic oscillator basis is applicable for light nuclei, but has to be replaced by better basis sets such as 
the Woods-Saxon plus $ls$ coupling model if heavier nuclei are considered. It is not the Coulomb part or the $N-N$ interaction contribution to the mean field $W^{\rm ext}( R)$ which makes the problems, 
but the antisymmetrization of fermionic wave functions and the Pauli blocking of the quartet with the core nucleus 
what makes the difficulties.

We investigate the independent particle case to understand the behavior of the quartet wave function 
inside the core nucleus. 
In particular we show that the effective potential $W( R)$ is nearly constant inside the core nucleus. 
We consider the model of free nucleons moving in a harmonic oscillator potential, see Sec. \ref{Harmonicoscillatormodel} and construct the 
effective potential $W^{\rm h.o.}( R)$ for the c.o.m. motion of a quartet. 
Furthermore, we consider the Thomas-Fermi rule and find $E_{\rm tunnel} > \mu_4$.

\subsection{Center of mass motion}
\label{Centerofmassmotion}

It is of interest to determine the c.o.m. motion of the quartet for the case of uncorrelated motion. We consider quartets formed by nucleons in zero angular momentum ($s$) orbitals. We use Jacobi-Moshinsky coordinates (\ref{Jacobi}) for the quartet nucleons.

\begin{figure}
\begin{center}
\includegraphics[%
  width=0.65\linewidth,
  keepaspectratio]{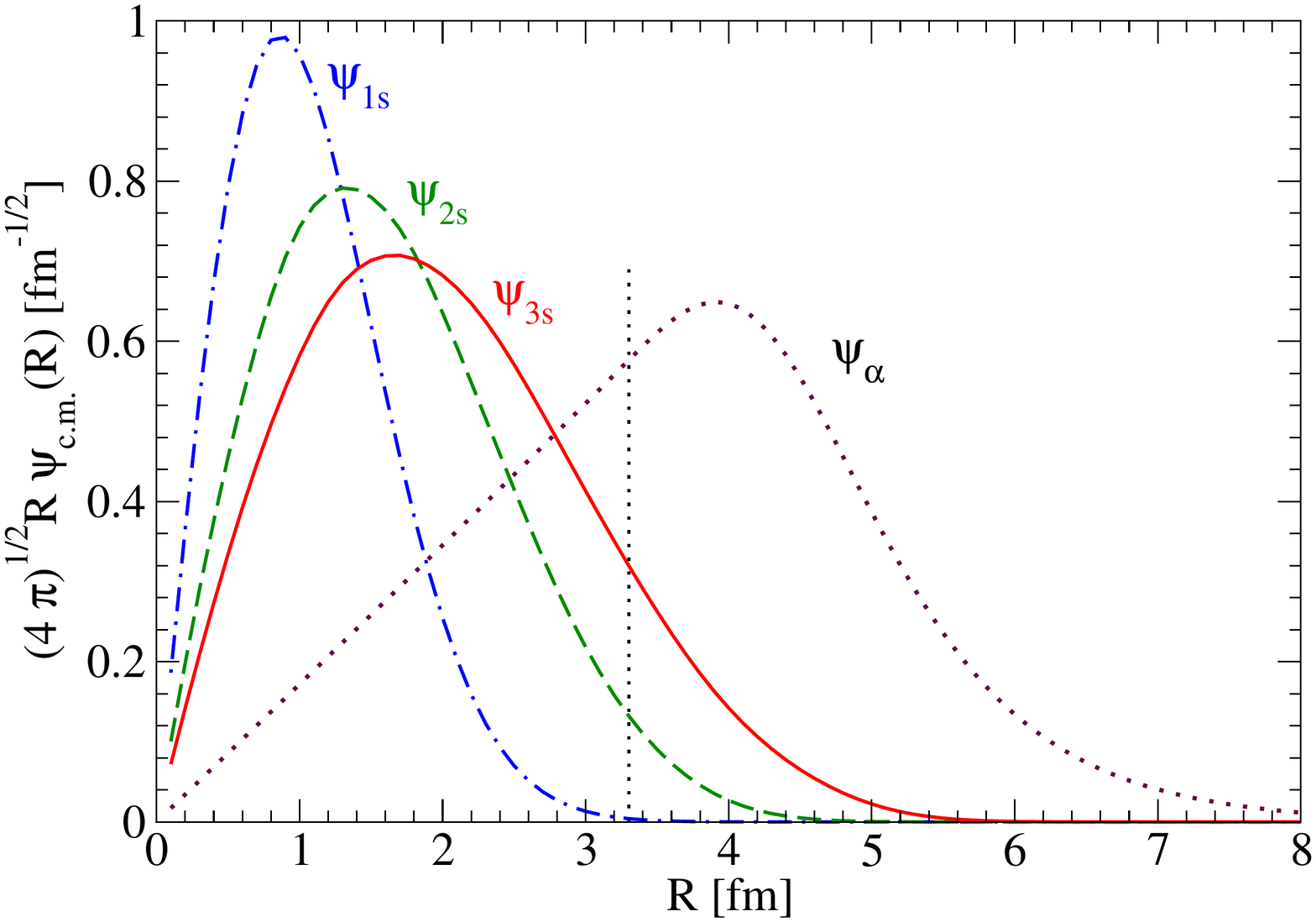}
\end{center}
\caption{Wave function for the c.o.m. motion of the non-interacting quartet. A prefactor $(4 \pi)^{1/2} R$ is introduced so that the integral over $r$ of the squared quantity is normalized to 1.
 The c.o.m. wave function of a  quartet consisting of 3s states (full line) is compared with the corresponding quantity for 2s states (dashed) and 1s states (dot-dashed). 
 For comparison, the result for the Thomas-Fermi calculation for the interacting quartet on top of $^{16}$O, Fig. \ref{fig:3}, is also shown (dotted). (Color figure online.)}
\label{fig:4}
\end{figure}

First we consider a quartet formed by uncorrelated nucleons in the lowest 1s state. 
The 1s orbital $\psi_{1s}( {\bf r})=\left(\frac{a}{\pi}\right)^{3/4} e^{-a r^2/2}$ gives the quartet wave function 
\begin{equation}
 \Phi_{1s^4}({\bf R, S, s, s}')=\psi_{1s}({\bf r}_{n,\uparrow})\psi_{1s}({\bf r}_{n,\downarrow})\psi_{1s}({\bf r}_{p,\uparrow})\psi_{1s}({\bf r}_{p,\downarrow})
\end{equation}
with the result
\begin{equation}
\label{Phi1s}
\psi_{1s^4}({\bf R})=\left[\int d^3Sd^3sd^3s'|\Phi_{1s^4}({\bf R, S, s, s}')|^2\right]^{1/2}=\left(\frac{4a}{\pi}\right)^{3/4} e^{-2 a R^2}
\end{equation}
A plot of $(4 \pi R^2)^{1/2} \psi_{1s^4}( R)$ is shown in Fig. \ref{fig:4}. The normalization\\ $\int_0^\infty 4 \pi R^2 \psi^2_{1s^4}( R) dR =1$ holds,
\begin{eqnarray}
\label{rho1s}
\varrho_{1s^4}^{\rm cm}( R)&=&|\psi_{1s^4}( R)|^2=\left(\frac{4a}{\pi}\right)^{3/2} 
e^{-4a R^2}.
\end{eqnarray}
(Note that the ground state which is a product of Gaussians can be considered as the independent motion in a harmonic oscillator potential,
but also as an $\alpha$-like cluster with the corresponding motion of the c.o.m. coordinate. 
This is also known from the THSR approach, see \cite{BBohr}.)

More interesting is the free motion of four nucleons in the harmonic oscillator potential on top of the  $^{16}$O like configuration which simulates the 
situation of $^{20}$Ne inside the core nucleus where correlations are suppressed because of the Pauli blocking. For this case also the c.o.m. motion is of interest.
 The calculation is performed according (\ref{Phi1s}) but using the wave function $\psi_{2s}( r)$ (\ref{psin2}). The integrals over the intrinsic Jacobi-Moshinsky coordinates can be performed.
 For the collective c.o.m. motion, the wave function follows after a cumbersome calculation  according to
\begin{eqnarray}
\label{rho2s}
&&\varrho_{2s^4}^{\rm cm}( R)=|\psi_{2s^4}( R)|^2=\left(\frac{a}{\pi}\right)^{3/2} e^{-4a R^2} \frac{1}{10616832} (24695649+14905152\, a R^2\nonumber \\&&
+354818304\, a^2R^4 -876834816\, a^3R^6
+1503289344\, a^4R^8-1261699072\, a^5R^{10} \nonumber\\ &&
+613416960\, a^6R^{12}-150994944\, a^7 R^{14}+16777216\, a^8R^{16})
\end{eqnarray}
Similar expressions for a mixed quartet  formed by two nucleons in $1s$ states and two nucleons in $2s$ states 
as well as a quartet formed by 4 nucleons in the $3 s$ state are given in App. \ref{app:1}.
 The corresponding plots are presented in Fig. \ref{fig:4}. 
 The intrinsic motion of the quartet is also given in App. \ref{app:1}.

\subsection{The effective potential for the c.o.m. motion}

We have constructed wave functions $\psi_{\nu}( R)=(\varrho_{\nu}^{\rm cm}( R) )^{1/2}$ for various contributing single-nucleon states.
The wave equation for the c.o.m. motion of the quartet has the form
\begin{equation}
\label{Sglcom}
-\frac{\hbar^2}{8 m} \nabla_R^2 \psi_{\nu}( R)+W_\nu^{\rm h.o.}( R) \psi_{\nu}( R)=E_\nu  \psi_{\nu}( R).
\end{equation}
Let us restrict to $s$ states ($l=0$) and introduce $u_\nu( R)= (4 \pi)^{1/2} R \psi_{\nu}( R)$, we have
\begin{equation}
\label{Sglcom1}
W_\nu^{\rm h.o.}( R)-E_\nu=\frac{\hbar^2}{8m} \frac{1}{u_\nu( R)} \frac{d^2}{dR^2} u_\nu( R).
\end{equation}
For the ground state $1s^4$ we have (with $a=m \omega/\hbar$) the result $W_{1s^4}^{\rm h.o.}( R)= 2 m \omega^2 R^2$ as expected,
each of the four nucleons feels the external (mean field) potential $m \omega^2 r^2/2$ so that the potential 
$W_{1s^4}^{\rm h.o.}( R)=4V^{\rm mf}( R)$ 
for the quartet follows. For the energy eigenvalue results $E_{1s^4} = 3 \hbar \omega/2$ in accordance for  the energy per degree of freedom of a harmonic oscillator. 

\begin{figure}
\begin{center}
\includegraphics[%
  width=0.65\linewidth,
  keepaspectratio]{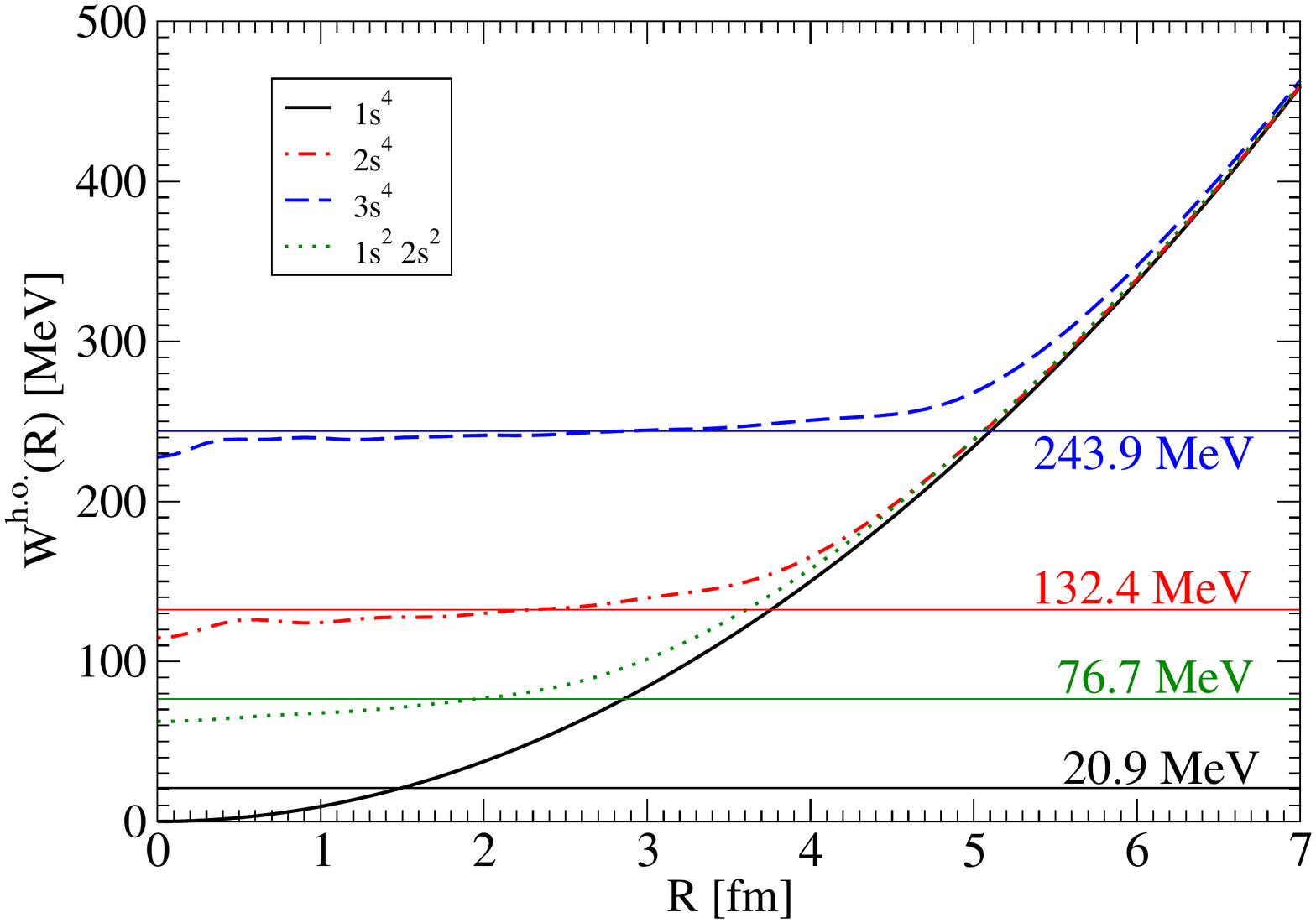}
\end{center}
\caption{Effective potential $ W_\nu^{\rm h.o.}( R)$ for the c.o.m. motion of the quartet. Ground state ($1s^4$), no medium, and on top of occupied states:
$1s^22s^2$,  $2s^4$, $3s^4$. Parameter values as for $^{16}$O in Sec. \ref{Harmonicoscillatormodel}. (Color figure online.)}
\label{fig:W_R}
\end{figure}

In the same way, also for the other states the effective potential $W_{\nu}^{\rm h.o.}( R)$
 has been calculated. The expressions are quite complex, therefore we give only a Figure
 \ref{fig:W_R}. The energy eigenvalues are $E_{2s^4} = 19 \hbar \omega/2,\,\,
E_{3s^4} = 35 \hbar \omega/2,\,\,E_{1s^22s^2} = 11 \hbar \omega/2$ also shown in Fig.
 \ref{fig:W_R}.
 We see that the condition $W^{\rm h.o.}( R) \approx$ const. is nearly fulfilled inside the core region, in particular for higher orbits. 
 In contrast to the Thomas-Fermi model where a kink occurs when $W^{\rm TF}( R)= \mu_4$, the transition is smooth.

\begin{figure}
\begin{center}
\includegraphics[%
  width=0.65\linewidth,
  keepaspectratio]{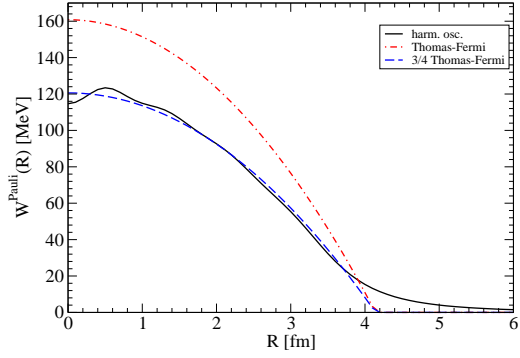}
\end{center}
\caption{Exchange potential $W^{\rm Pauli}_{2s^4}( R)$ for the c.o.m. motion of the quartet. State   $2s^4$ compared with
the value $4\, E_F( R)$ of the Thomas-Fermi model; for comparison also $3\, E_F( R)$ has been shown. 
Parameter values as for $^{16}$O in Sec. \ref{Harmonicoscillatormodel}. (Color figure online.)}
\label{Paulipot16O}
\end{figure}

We can also consider the exchange potential $W^{\rm Pauli}_{2s^4}( R) = W^{\rm h.o.}_{2s^4}( R)-4V^{\rm mf}( R)$ shown in Fig. \ref{Paulipot16O}.
The Thomas-Fermi model gives
$4 E_F( R)= 4 [40.2048\,\, {\rm MeV} -V^{\rm mf}( R)]$ also shown there, with $V^{\rm mf}( R) = 2.3429\,\, {\rm MeV\,\, fm}^{-2} R^2$ 
for $^{16}$O.
The values are rather large. Better correspondence is obtained for $3 E_F( R)$.

\subsection{Overlap with the alpha}

The $\alpha$ particle is given in Gaussian approximation by the intrinsic wave function
\begin{equation}
\label{rhoalpha}
\varphi^{\rm intr}_{\alpha}({\bf S,s,s'})=\left(\frac{a}{\pi}\right)^{9/4} 2^{-3/2}e^{-\frac{a}{4} (2 {\bf S}^2+{\bf s}^2+{\bf s'}^2)}.
\end{equation}
It is normalized,
\begin{equation}
 \int d^3S\,\,d^3\!\!s\,\,d^3\!\!s'\left[\varphi^{\rm intr}_{\alpha}({\bf S,s,s'})\right]^2=1.
\end{equation}
The point rms radius follows as
\begin{eqnarray}
 \int d^3\!S\,\,d^3\!\!s\,\,d^3\!\!s'&[&({\bf r}_1-{\bf R})^2+({\bf r}_2-{\bf R})^2+({\bf r}_3-{\bf R})^2
+({\bf r}_4-{\bf R})^2]/4 \left[\varphi^{\rm intr}_{\alpha}({\bf S,s,s'})\right]^2
\nonumber\\&&={\rm rms}_\alpha^2=\frac{9}{8a}.
\end{eqnarray}
With ${\rm rms}_\alpha=1.45$ fm follows $a_\alpha=0.535077$ fm$^{-2}$.

We consider the uncorrelated quartet states of the $^{16}$O nucleus which are calculated with $a_{\rm O}=0.33619$ fm$^{-2}$.
In the $1s$ state we calculate the overlap with the $\alpha$ wave function
\begin{eqnarray}
\langle\varphi^{\rm intr}_{\alpha}|\varphi^{\rm intr}_{1s^4} \rangle&=& \int d^3S\,\,d^3\!\!s\,\,d^3\!\!s'
{\varphi^{\rm intr}_{\alpha}}^*({\bf S,s,s'}) \varphi^{\rm intr}_{1s^4}({\bf S,s,s'})
\nonumber\\&=& \frac{2^{9/2} a_\alpha^{9/4} a_{\rm O}^{9/4}}{ (a_\alpha+a_{\rm O})^{9/2}} = 0.886557.
\end{eqnarray}
We can consider $|\langle \varphi_{{\rm intr.},\alpha}|\varphi_{{\rm intr.},1s^4} \rangle|^2=0.785983$ as probability to find in this $1s$ quartet of $^{16}$O the $\alpha$ particle. 
It is clear that the localization in an external potential looks like the formation of correlation, 
and for a suitable harmonic oscillator potential we would obtain the $1s$ quartet identical with the intrinsic wave function of the $\alpha$ particle. 
Because the mean-field potential of the $^{16}$O nucleus (or, more essential, the $^{20}$Ne nucleus) is more extended, the intrinsic $1s$ quartet wave function gives a smaller overlap. 

More interesting is the overlap of the alpha with the localized states in the $2s$ orbital. With the intrinsic wave function (\ref{2sintr}), we calculate
\begin{eqnarray}
&& \langle \varphi_{{\rm intr.},\alpha}|\varphi_{{\rm intr.},2s^4} \rangle(R)= \int d^3\!S\,\,d^3\!\!s\,\,d^3\!\!s'
 \varphi_{{\rm intr.},\alpha}^*({\bf S,s,s'}) \varphi_{{\rm intr.},2s^4}({\bf S,s,s';R})\nonumber\\ 
&& = \frac{1024 a^{9/4}a_\alpha^{9/4}}{3^{3/2} (a+a_\alpha)^{17/2}}\nonumber\\ 
&&\times (9 (41 a^4 -152 a^3 a_\alpha+360 a^2 a_\alpha^2-288 a a_\alpha^3+144 a_\alpha^4)
\nonumber\\ 
&&-144 a (a+ a_\alpha) (a^3+22 a^2 a_\alpha-12 a^2 a_\alpha^2+24 a_\alpha^3) R^2\nonumber\\ 
&&+288 a^2 (a+a_\alpha)^2 (5a^2+4 a a_\alpha+12 a_\alpha^2) R^4
\nonumber\\ 
&&-768 a^3 (a+a_\alpha)^3 (a+2 a_\alpha) R^6+256 a^4 (a+a_\alpha)^4 R^8)
\nonumber \\ &&
\times(24695649+14905152\, a R^2+354818304\, a^2R^4 -876834816\, a^3R^6
\nonumber\\ 
&&+1503289344\, a^4R^8-1261699072\, a^5R^{10}+613416960\, a^6R^{12}\nonumber\\ 
&&-150994944\, a^7 R^{14}+16777216\, a^8R^{16})^{-1/2}\,.
\end{eqnarray}
The probability to find the $\alpha$ particle in the $2s$ state is 
\begin{equation}
\int_0^\infty dR\,\, 4 \pi R^2 \varrho_{{\rm c.o.m.}, 2s^4}( R) |\langle \varphi_{{\rm intr.},\alpha}|\varphi_{{\rm intr.},2s^4} \rangle(R)|^2 = 0.00115899
 \end{equation}
 for the values $a=a_{\rm O}= 0.33619$ fm$^{-2}$ and $a_\alpha=0.535077$ fm$^{-2}$. In the region where the wave function of the quartet is approximated by a product of nearly free single-particle orbitals, the preformation of an $\alpha$ particle is very low.

\section{Quartetting}
\label{Quartetting}

To describe quartetting, we have to go beyond the uncorrelated motion of nucleons in a mean-field potential as considered in the harmonic oscillator model. The nucleon-nucleon interaction within the quartet leads to the formation of correlations. 
The wave equation (\ref{15}) describes in the 
zero density limit the $\alpha$ particle as state with lowest energy. After separation of the c.o.m. motion with energy $\hbar^2 P^2/8m$,
then the intrinsic part gives the contribution to the effective potential $W^{\rm intr}( R)=-28.3$ MeV. 
Taking into account the blocking terms, this contribution is changed, and  we obtain with Eq. (\ref{WPauli}) the result
\begin{equation}
\label{WintrR}
 W^{\rm intr}( R)=-28.3\,{\rm  MeV} +W^{\rm Pauli}[n_B(R )].
\end{equation}
Note that the expression (\ref{WPauli}) follows from homogeneous matter and is used here as local density approximation.
As discussed in \cite{Po}, Pauli blocking is non-local and cannot be rigorously represented by a local potential.

\begin{figure}
\begin{center}
\includegraphics[%
  width=0.65\linewidth,
  keepaspectratio]{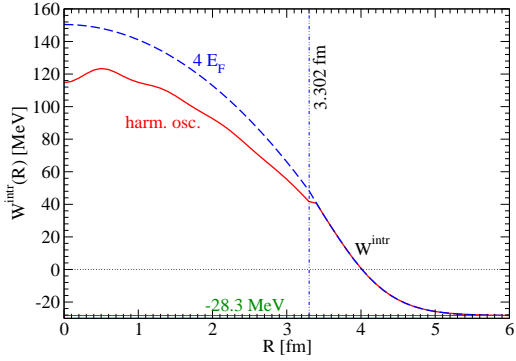}
\end{center}
\caption{Intrinsic potential $W^{\rm intr}_{2s^4}( R)$ for the c.o.m. motion of the quartet. For  $R \le r_{\rm crit}=3.302$ fm
the harmonic oscillator result is compared with
the position of the Fermi energy $4 E_F( R) = \mu_4- W^{\rm mf}( R)$ within the Thomas-Fermi model. 
The value $\mu_4= 150.402$ MeV was adapted to the h.o. density at $r_{\rm crit}$. For $R \ge r_{\rm crit}$,
the local density approximation (\ref{WintrR}) is taken.
Parameter values as for $^{16}$O in Sec. \ref{Harmonicoscillatormodel}. (Color figure online.)}
\label{WintPauli}
\end{figure}

As soon as the bound state disappears if the critical density is reached, the uncorrelated  intrinsic motion sets in.
We find $W^{\rm intr}( R)\approx 4 E_F( R) $ for $R \le r_{\rm crit}$.
The intrinsic potential $W^{\rm intr}_{2s^4}( R)$ for the c.o.m. motion of the quartet formed 
by the $2s^4$ state is shown in Fig. \ref{WintPauli}. 
The small misfit at $r_{\rm crit}$ is caused by reason that for $\mu_4$ the harmonic oscillator density 
$n_B( R)$ has been fitted which is not identical with the density obtained from the Thomas-Fermi model. 

We conclude that a nearly constant effective potential  $W( R)$ inside the core nucleus, see Fig. \ref{fig:W_R}, can be understood. 
Shell model calculations will improve the detailed form of this potential, as already seen using the harmonic oscillator basis.

Because the harmonic oscillator potential $W^{\rm mf}( R)$ is rather steep, no pocket is formed. Realistic mean-field potentials are weakening at large $R$ so that a pocket is formed as shown in Refs \cite{Po,Xu,Xu1}.

\begin{figure}
\begin{center}
\includegraphics[%
  width=0.65\linewidth,
  keepaspectratio]{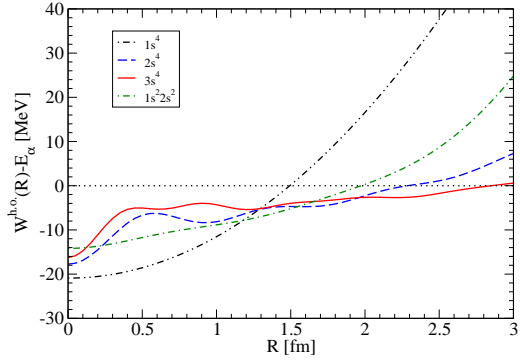}
\end{center}
\caption{Difference between the effective potential $ W^{\rm h.o.}( R)$ for the c.o.m. motion of the quartet 
and the corresponding energy eigenvalue. States $1s^4$, $1s^22s^2$,  $2s^4$, $3s^4$. 
Parameter values as for $^{16}$O in Sec. \ref{Harmonicoscillatormodel}.  (Color figure online.)}
\label{DelW3s}
\end{figure}

It would be of interest to reproduce the value $\mu_4$ of the Thomas-Fermi model in the harmonic oscillator 
shell model. Instead of the constant value $\mu_4$, the effective potential $W( R)$ is depending on $R$. 
We show examples in Fig. \ref{DelW3s}. With increasing quantum number $n$, 
the effective potential becomes flat and closer to the energy eigenvalue as expected from the Thomas-Fermi model.

Within the Thomas-Fermi model, a significant shell effect has been obtained for the difference 
$\Delta = E_\alpha-\mu_4$.
It is challenging to reproduce this effect within the present calculations. For this we consider a quartet 
of two nucleons in the $2s$ state and two in the $1s$ state.
Two reasons may be considered to contribute to the shell effect: (i) The difference $\Delta$ becomes larger 
if the lower shell is used to form a quartet, it is stronger bound. (ii) The intrinsic potential $W^{\rm intr.}( R)$ 
 outside the critical radius $r_{\rm crit}$ is less relevant because it is determined by 
the total nucleon density (which changes smoothly if the mass number of the core nucleus is changed), 
but the extension of the 
c.o.m. wave function is strongly reduced so that the part in the surface region where $\alpha$ 
clusters may exist, is diminished. The explanation of the shell effect  needs further work.

Whereas this effective potential is smooth, the intrinsic wave function changes abruptly at $r_{\rm crit}$. 
The $\alpha$-like wave function changes to a nearly uncorrelated product of single-nucleon states.

\section{Consequences for quartetting in nuclei}

\subsection{Consequences for  the $^{212}$Po calculation}

The mean-field potential $V^{\rm mf}$ has been obtained from double folding of Coulomb and 
nucleon-nucleon interaction, see \cite{Po,Xu,Xu1} where more details are given. 
In particular, within the Thomas-Fermi approach, the parameter values
for $c,d$ are given. Of interest would be the use of shell model wave functions for the $^{208}$Pb core nucleus. With an appropriate 
Woods-Saxon + $ls$ mean-field potential, the fit of the parameter $\mu_4$ in the Thomas-Fermi model may be avoided.

Recently \cite{Xu1} the series of Po isotopes has been considered, and the signatures of the magic numbers have been found.
As discussed in Sec. \ref{Quartetting} considering quartetting of two $2s$ nucleons (e.g. neutrons) with two $1s$ nucleons (e.g. protons),
consequences for the $\alpha$ preformation factor are expected as observed from experiments.

\subsection{Consequences for the $^{20}$Ne calculation: relaxing the Thomas-Fermi rule}
\label{Consequences20Ne}

The parametrization of the mean-field potential for a quartet on top of the $^{16}$O core nucleus has been given in Sec. \ref{mf}.
The condition  $E_\alpha = \mu_4$ is a consequence of of the Thomas-Fermi model valid for infinite matter: an additional nucleon with given spin and isospin can be introduced at the corresponding chemical potential $\mu_{\sigma, \tau}$. This coincides at zero temperature with the corresponding Fermi energy (plus the potential energy). For finite system such as nuclei, 
the energy levels of the single-nucleon states are discrete. When we add a nucleon to the core nucleus where all single-nucleon states below a certain energy are occupied, the next free single-nucleon state which is free has a distance from the chemical potential. This means, that under these considerations the quartet cannot be introduced at $\mu_4$ 
but at a higher value $E_\alpha > \mu_4$ which is now a new parameter. This aspect has been worked out already  in \cite{Xu}. 
We do the same here for $^{20}$Ne.

The $\alpha$-decay energy $Q_\alpha$ was introduced as difference of the binding energy of the mother nucleus ($^{212}$Po)
 and the binding energies of the daughter nuclei (Pb and $\alpha$). Similarly  we have -4.73 MeV so that the energy eigenvalue 
 of the Schr\"odinger equation $E^0_\alpha-Q_\alpha=-28.3-4.73$  MeV=-33.03 MeV.
 As second condition we used the results for $^{212}$Po. If $d=3415.56$ MeV remains the same, the given energy eigenvalue 
 of the Schr\"odinger equation is reproduced with $c=10623$ MeV. Then, the value $\mu_4=-32.388$ MeV and $P_\alpha=0.72$
 follow. If we take from Po $c=11032$, we find $d=3513.46$ and $\mu_4=-32.12$ MeV and $P_\alpha=0.74$ result.
 
 We reproduce in both cases a large preformation factor $P_\alpha$. In contrast to the Thomas-Fermi model, 
 the condition  $E_\alpha = \mu_4$ is not valid in general. 
 The value of $\mu_4$ is not below $E_\alpha$ as expected from the shell model 
 consideration, but $E_\alpha < \mu_4$. This means that for the core nucleus it is energetically favored to form correlated quartets
 instead to stay in uncorrelated single- nucleon (shell model) states. This will be seen from the THSR calculations where
 the core nucleus $^{16}$O shows also $\alpha$-like correlations.

If the $^{20}$Ne is described by the uncorrelated harmonic oscillator shell model, the energy of the $2s$ state and the corresponding quartet 
are too high. Correlations, in particular $\alpha$-like correlations, will reduce the energy. 
At the same time, the relation between the rms radius and the parameter $\hbar \omega$ of the potential will change.
The wave function for the c.o.m. motion is smooth, the increase at small values of $R$ goes down when higher orbitals $ns$ are considered. This tendency can be extrapolated to $^{212}$Po \cite{Po,Xu}.
The intrinsic rms radius for the $2s$ orbit is not very different from the $\alpha$ particle, 
but the overlap with the $\alpha$ particle is small as expected for the uncorrelated motion. In particular it is small for $R < 3$ fm,
but increases in the outer region. The preformation probability is very small, but not identical to zero because the localized states always give 
a contribution which looks like a correlation, also within a Hartree-Fock approach. Probably the strong reduction of the preformation probability
is because of the antisymmetrization of the $2s$ quartet with respect to the $^{16}$O core nucleus.

To improve the harmonic oscillator model, correlations must be implemented as shown in the next Section. 
A class of wave functions will be considered which, in contrast to the shell model approach, allow for $\alpha$-like correlations
which we denote as quartet states, with a c.o.m. motion different from the intrinsic motion. 
This wave function can be optimized looking for the minimum of energy.\\

\section{Conclusions, comparison with the THSR model}

We investigated the properties of an $\alpha$-like quartet 
moving on the top of a core nucleus. 
The effective Schr{\"o}dinger equations for the c. o. m. motion in the 
mean-field potential of the core nucleus as well as for the intrinsic motion
are considered. 
In particular, for the c.o.m. motion of the quartet an effective potential $W( R)$ 
has been given, which shows a pocket structure near the surface of the nucleus 
what is of relevance for the preformation of $\alpha$ particles. 
A new aspect is the behavior of  $W( R)$ within the core nucleus, 
i.e. for $R \le r_{\rm crit}$ where the bound state is dissolved
because of Pauli blocking. 
In contrast to former investigations which assume an increase of this
effective potential with decreasing $R$, within a Thomas-Fermi approach it can argued 
that $W(R)=\mu_4$ remains constant in this region \cite{Po,Xu}, see Fig. \ref{fig:2}. 
In the present work we show that such a behavior can also be derived from
a shell model approach. Our main result is the treatment of the exchange part 
of the effective potential. Using the harmonic oscillator basis, we performed model calculations 
to show that $W(R)$ remains nearly constant. Note that the harmonic oscillator wave functions are considered as
a model to investigate the behavior of the effective potential $W( R)$ within 
the core region. In the outer region of the nucleus, the behavior 
of the wave function has to be derived from the nuclear mean-field potential,
for instance a   Woods-Saxon potential.

The Thomas-Fermi model is quite simple and gives a mean-field description for the quartet.
Because it is a local-density approach, it is not appropriate to describe shell effects 
as observed, e.g., for the Po isotopes, see Ref. \cite{Xu1}. In particular, the 
Thomas-Fermi rule $E_{\rm tunnel}=\mu_4$ (Sec. \ref{TFrule}) 
is too restrictive, and a gap has been introduced empirically to obtain 
realistic results for the $\alpha$-decay life times of Po isotopes \cite{Po,Xu}.

This difference between the chemical potential of the Thomas-Fermi model 
and the next higher energy level can be obtained introducing a shell model 
for the core nucleus. Then, the Pauli blocking is no longer simple as for homogeneous matter.
Analytical calculations are presented for a harmonic oscillator wave functions basis. Future calculations should work with a more realistic mean field as
common in shell-model calculations, 
based on a nucleon-nucleon interaction including exchange terms.
It would be of interest to use 
actual shell model states, obtained from a Woods-Saxon + $ls$ potential. 
The main issue is to obtain as effective potential $W( R)$ for the c.o.m. motion 
which describes the formation of a pocket and a nearly constant potential inside the core nucleus.
A problem to be solved in future investigations is the treatment of partially filled shells,
when spherical symmetry can not longer be assumed. 

The direct comparison with the THSR approach which treats the quartets self-consistently needs further work. If the mean-field approach is no longer possible, the full antisymmetrization of the many-body wave function is very challenging. Until now, the THSR
approach provides us with such a self-consistent treatment of all nucleons. 
A variational principle with Gaussian wave functions has been used, and nuclei with $A \le 20$
have been treated this way. 

The comparison of the quartetting wave function approach 
with the THSR approach may answer the question whether quartetting is also relevant for the core nucleus, in contrast to the core nucleus shell model considered here.
It is known \cite{THSR} that $\alpha$-like correlations are also present 
in the ground state of the $^{16}$O core nucleus. Within THSR calculations, the minimum of the energy functional 
has been found for Gaussians
with different width $b, B$ for the intrinsic cluster wave function 
and the c.o.m. wave function, respectively, see  Fig. 2 of \cite{THSR}. 
This indicates that the assumption of a shell model of independent single-nucleon orbits is not fully justified for the $^{16}$O core nucleus.
Only for $b=B$ a pure shell model is obtained \cite{BBohr}. 
Note that in the THSR calculations \cite{Bo3,Bo,Bo2} for $^{20}$Ne also 
a Gaussian shell model state (\ref{ho}) was taken for the $^{16}$O core nucleus.

\appendix

\section{Calculations with harmonic oscillator wave functions}
\label{app:1}

We present some results of calculations with harmonic oscillator wave functions, the corresponding figures are shown in the main text.
In addition to (\ref{psin1}), we use
\begin{eqnarray}
\label{psin2}
&& \psi_{1d}( r)=\left(\frac{a}{\pi}\right)^{3/4} e^{-a r^2/2} \left(\frac{4 a^2}{15}\right)^{1/2} r^2\,\,Y_{2m},\nonumber\\
&&  \psi_{1f}( r)=\left(\frac{a}{\pi}\right)^{3/4} e^{-a r^2/2}\left(\frac{8 a^3}{105}\right)^{1/2}r^3\,\,Y_{3m},\nonumber\\
&&  \psi_{2s}( r)=\left(\frac{a}{\pi}\right)^{3/4} e^{-a r^2/2} \left( \frac{2}{3}\right)^{1/2}\left(a r^2-\frac{3}{2}\right),\nonumber\\
&&  \psi_{2p}( r)=\left(\frac{a}{\pi}\right)^{3/4} e^{-a r^2/2}\left(\frac{4 a}{15}\right)^{1/2}r 
\left(a r^2-\frac{5}{2}\right)\,\,Y_{1m},\nonumber\\
&&  \psi_{3s}( r)=\left(\frac{a}{\pi}\right)^{3/4} e^{-a r^2/2} \left( \frac{2}{15}\right)^{1/2}\left(a^2 r^4- 5 a r^2+\frac{15}{4}\right).
\end{eqnarray}

We give some results for the c.o.m. motion $\psi_\nu( R)$ of the quartet.
For a mixed state where the quartet is formed by two nucleons in  $1s$ states and two nucleons in $2s$ states we have
\begin{eqnarray}
\label{rho1s2s}
&&\varrho_{1s^22s^2}^{\rm cm}( R)=|\psi_{1s^22s^2}( R)|^2=\left(\frac{a}{\pi}\right)^{3/2} e^{-4a R^2} \frac{1}{1152}
\nonumber \\ && \times
(4305+32\, a R^2 (173+4\, a R^2 (77+16\, a R^2 (-1+2 a R^2))))
\end{eqnarray}

We add the result for the quartet of nucleons in the 3s state:
\begin{eqnarray}
\label{rho3s}
&&\varrho_{3s^4}^{\rm cm}( R)=|\psi_{3s}( R)|^2=\left(\frac{a}{\pi}\right)^{3/2} e^{-4a R^2} \frac{1}{111325552312320000} \nonumber\\ &&
\times (132383603722601025+118077233897001600\, a R^2
 \nonumber\\ &&
+3512361784297996800\, a^2 R^4 -21123743680445767680\, a^3 R^6 \nonumber\\ &&
 +93657803058195578880\, a^4 R^8
-248906350014504632320\, a^5 R^{10} \nonumber\\ &&
+436533099609204981760\, a^6 R^{12}-523813234956493127680\, a^7 R^{14}
 \nonumber\\ &&
+443662383082136141824\, a^8 R^{16}-269879474498698215424\,a^9 R^{18} \nonumber\\ &&
+ 118989746552373772288 \, a^{10} R^{20}
-38020752685630750720\, a^{11} R^{22} \nonumber\\ &&
+8714612338642124800\,a^{12} R^{24}- 1397461686717251584\,a^{13} R^{26}
 \nonumber\\ &&
+ 149216922028736512 \, a^{14} R^{28}-9570149208162304\, a^{15} R^{30} \nonumber\\ &&
+ 281474976710656\,a^{16} R^{32})
\end{eqnarray}

We give some results for the intrinsic motion of the quartet.
After separating the c.o.m. motion $\psi_\nu( R)$ of the quartet, the intrinsic motion remains. 
Using Jacobi-Moshinsky coordinates we have
\begin{equation}
\label{1sintr}
 \varphi_{1s^4}^{\rm intr}({\bf S},{\bf s}_1,{\bf s}_2;{\bf R})=
\left(\frac{a}{\pi}\right)^{9/4} \frac{1}{2^{3/2}}e^{-\frac{a}{4}(2 {\bf S}^2+{\bf s}_1^2+{\bf s}_2^2)},
\end{equation}
no dependence on $R$ appears. The normalization $\int d^3S\,d^3s_1\,d^3 s_2 \,| \varphi_{1s^4}^{\rm intr}({\bf S},{\bf s}_1,{\bf s}_2;{\bf R})|^2 =1$ holds.

For the $2s$ state we have
\begin{equation}
 \varphi_{2s^4}^{\rm intr}({\bf S},{\bf s}_1,{\bf s}_2;{\bf R})=\psi_{2s}({\bf r}_1)\psi_{2s}({\bf r}_2)\psi_{2s}({\bf r}_3)\psi_{2s}({\bf r}_4)
\frac{1}{\psi_{2s^4}( R)}
\end{equation}
so that
\begin{eqnarray}
\label{2sintr}
&& \varphi_{2s^4}^{\rm intr}({\bf S},{\bf s}_1,{\bf s}_2;{\bf R})=
\left(\frac{a}{\pi}\right)^{9/4} 2^{5/2}e^{-\frac{a}{4}(2 {\bf S}^2+{\bf s}_1^2+{\bf s}_2^2)}
\nonumber \\ &&\times [6+a(2{\bf R}+{\bf S}+{\bf s}_1)^2][6+a(2{\bf R}+{\bf S}-{\bf s}_1)^2]
\nonumber \\ &&\times [6+a(2{\bf R}-{\bf S}+{\bf s}_2)^2][6+a(2{\bf R}-{\bf S}-{\bf s}_2)^2]
\nonumber \\ &&\times (24695649+14905152\, a R^2+354818304\, a^2R^4 -876834816\, a^3R^6\nonumber \\ &&+1503289344\, a^4R^8
-1261699072\, a^5R^{10}+613416960\, a^6R^{12} \nonumber\\ &&
-150994944\, a^7 R^{14}+16777216\, a^8R^{16})^{-1/2}\,.
\end{eqnarray}
The normalization $\int d^3S\,d^3s_1\,d^3 s_2 \,| \varphi_{2s^4}^{\rm intr}({\bf S},{\bf s}_1,{\bf s}_2;{\bf R})|^2 =1$ holds.
In the limit $R\to \infty$ it coincides with $\varphi_{1s^4}^{\rm intr}({\bf S},{\bf s}_1,{\bf s}_2;{\bf R})$, Eq. (\ref{1sintr}).

We introduce in addition to the quartet c.o.m. position $\bf R$ also the distance $\bf r'=r-R$ of a nucleon. We give the density 
$\varrho_{n,s}^{\rm intr.}( {\bf r'; R})$ so that the nucleon density  at position $\bf r$ of the quartet at position $\bf R$ follows as $4 \varrho_{n,s}^{\rm intr.}( {\bf r'; R}) \varrho_{n,s}^{\rm cm}( R)$ (4 nucleons contribute equally to the density).

Of interest is the intrinsic wave function of the uncorrelated quartet in the 2$s$ state. For $R=0$, the density distribution of a nucleon (factor 4 for all nucleons of the quartet) at $\bf r$ is 
\begin{eqnarray}
 &&\varrho_{2s^4}^{\rm intr}(r; 0)=\frac{ 4 \times 131072}{ 2743961\times 3^{31/2}} \frac{a^{3/2}}{\pi^{3/2}} e^{-4 a\,r^2/3} (3-2 a r^2)^2  \nonumber\\ && \times
(40999689-10261404\,a\, r^2+7881948\,a^2\, r^4
-1267488\,a^3\,r^6 \nonumber\\ &&
+109296\,a^4\, r^8-4032\,a^5\,r^{10}+64\,a^6\, r^{12})
\end{eqnarray}
The integral $4 \pi \int r^2 d r\varrho_{2s^4}^{\rm intr}(r;0)$ is normalized to 1.

More general for arbitrary $\bf R$, we take it in $z$ direction, and $\bf r$ in the $x-z$ plane. Replace $\bf S=2 r-2R-s_1$ and a factor 8 replacing in the $\delta$ function $\bf S$ by $\bf r$. 
In the $1s$ orbital we have 
\begin{equation}
\label{rhoint1s}
\varrho_{1s^4}^{\rm intr.}( {\bf R, r'})=\left(\frac{4 a}{3 \pi}\right)^{3/2}e^{-4 a r'^2/3}
\end{equation}
independent on the distance $R$.

The 2s yields: (${\bf R \cdot r'}=R \,r'\, z$;  $a$ is dropped)
\begin{eqnarray}
&&\varrho_{2s^4}^{\rm intr.}( {\bf r'; R})=\left(\frac{4 a}{3 \pi}\right)^{3/2}e^{-4 a r'^2/3}\frac{65536}{531441} 
(3-2 (R^2+ 2 R r'z+r'^2))^2
\nonumber\\ &&
\times (40999689 - 10261404 r'^2 + 7881948 r'^4 - 1267488 r'^6 
\nonumber\\ &&
+109296 r'^8 - 4032 r'^{10} + 64 r'^{12} - 92352636 R^2 + 141875064 r'^2 R^2 \nonumber\\ &&
- 34222176 r'^4 R^2 + 3934656 r'^6 R^2 - 181440 r'^8 R^2 + 3456 r'^{10} R^2 \nonumber\\ &&
+ 638437788 R^4 - 307999584 r'^2 R^4 + 53117856 r'^4 R^4 - 3265920 r'^6 R^4 
\nonumber\\ &&
+77760 r'^8 R^4 - 923998752 R^6 + 318707136 r'^2 R^6 - 29393280 r'^4 R^6 
\nonumber\\ &&
+ 933120 r'^6 R^6 + 717091056 R^8 - 132269760 r'^2 R^8 + 6298560 r'^4 R^8 
\nonumber\\ &&
- 238085568 R^{10} + 22674816 r'^2 R^{10} + 34012224 R^{12} + 61568424 r' R z 
\nonumber\\ &&
-94583376 r'^3 R z + 22814784 r'^5 R z - 2623104 r'^7 R z + 120960 r'^9 R z 
\nonumber\\ &&
- 2304 r'^{11} R z - 851250384 r' R^3 z + 410666112 r'^3 R^3 z - 70823808 r'^5 R^3 z 
\nonumber\\ &&
+ 4354560 r'^7 R^3 z - 103680 r'^9 R^3 z + 1847997504 r' R^5 z - 637414272 r'^3 R^5 z 
\nonumber\\ &&
+58786560 r'^5 R^5 z - 1866240 r'^7 R^5 z -1912242816 r' R^7 z + 352719360 r'^3 R^7 z 
\nonumber\\ &&
-16796160 r'^5 R^7 z + 793618560 r' R^9 z -75582720 r'^3 R^9 z - 136048896 r' R^{11} z 
\nonumber\\ &&
+283750128 r'^2 R^2 z^2 - 136888704 r'^4 R^2 z^2 +23607936 r'^6 R^2 z^2 - 1451520 r'^8 R^2 z^2 
\nonumber\\ &&
+34560 r'^{10} R^2 z^2 - 1231998336 r'^2 R^4 z^2 +424942848 r'^4 R^4 z^2 - 39191040 r'^6 R^4 z^2 
\nonumber\\ &&
+1244160 r'^8 R^4 z^2 + 1912242816 r'^2 R^6 z^2 -352719360 r'^4 R^6 z^2 
\nonumber\\ &&
 + 16796160 r'^6 R^6 z^2 -1058158080 r'^2 R^8 z^2 + 100776960 r'^4 R^8 z^2
\nonumber\\ &&
 +226748160 r'^2 R^{10} z^2 + 273777408 r'^3 R^3 z^3 -94431744 r'^5 R^3 z^3
\nonumber\\ &&
 + 8709120 r'^7 R^3 z^3 -276480 r'^9 R^3 z^3 - 849885696 r'^3 R^5 z^3 
+156764160 r'^5 R^5 z^3
\nonumber\\ &&
 - 7464960 r'^7 R^5 z^3 +705438720 r'^3 R^7 z^3 - 67184640 r'^5 R^7 z^3 
-201553920 r'^3 R^9 z^3
\nonumber\\ &&
 + 141647616 r'^4 R^4 z^4 -26127360 r'^6 R^4 z^4 + 1244160 r'^8 R^4 z^4 
-235146240 r'^4 R^6 z^4 
\nonumber\\ &&
+ 22394880 r'^6 R^6 z^4 +100776960 r'^4 R^8 z^4 + 31352832 r'^5 R^5 z^5 
\nonumber\\ &&
-2985984 r'^7 R^5 z^5 - 26873856 r'^5 R^7 z^5 +2985984 r'^6 R^6 z^6)\nonumber\\ &&
\times \left(24695649+14905152\, a R^2+354818304\, a^2R^4 -876834816\, a^3R^6
\right. \nonumber\\ &&
\left.+1503289344\, a^4R^8-1261699072\, a^5R^{10}+613416960\, a^6R^{12}\right. \nonumber\\ &&
\left.-150994944\, a^7 R^{14}+16777216\, a^8R^{16}\right)^{-1}.
\end{eqnarray}

Performing the angular average ($1/2 \int_{-1}^1 dz$) we obtain (decomposition in spherical harmonics)
\begin{eqnarray}
&&\varrho_{2s^4}^{\rm intr.,0}( r';R)=\left(\frac{4 a}{3 \pi}\right)^{3/2}e^{-4 a r'^2/3}\frac{10616832}{43046721} \nonumber\\ &&
\times [512 r'^{16} + 1024/3 r'^{14} (-99 + 284 R^2) + 512 r'^{12} (1899 - 8718 R^2 + 6892 R^4) 
\nonumber\\ &&
+ 2304/7 r'^{10} (-38997 + 254492 R^2 - 318108 R^4 + 102544 R^6) 
\nonumber\\ &&
+ 1728/35 r'^8 (1933155 - 13878060 R^2 + 23760968 R^4 - 12187120 R^6
\nonumber\\ &&
 + 1750320 R^8) 
+ 15552/35 r'^6 (-661815 + 6497820 R^2 - 13251224 R^4
\nonumber\\ &&
 +  9828192 R^6 
- 3080880 R^8 + 411840 R^{10}) 
+ 39366 (3 - 2 R^2)^2 
\nonumber\\ &&
\times (2083 - 4692 R^2 + 32436 R^4
- 46944 R^6 + 36432 R^8 - 12096 R^{10} + 1728 R^{12}) 
\nonumber\\ &&
+ 69984/5 r'^4 (51165 - 438950 R^2 + 1155660 R^4 - 1306768 R^6 + 770352 R^8
\nonumber\\ &&
 - 236640 R^{10} + 33600 R^{12}) 
+104976 r'^2 (-11133 + 65036 R^2 - 211492 R^4
\nonumber\\ &&
 + 351536 R^6
 -275952 R^8 + 121920 R^{10} - 26304 R^{12} + 2304 R^{14})]\nonumber\\ &&
\times \left(24695649+14905152\, a R^2+354818304\, a^2R^4 -876834816\, a^3R^6
\right. \nonumber\\ &&
\left.+1503289344\, a^4R^8-1261699072\, a^5R^{10}+613416960\, a^6R^{12}
\right. \nonumber\\ &&
\left.-150994944\, a^7 R^{14}+16777216\, a^8R^{16}\right)^{-1}.
\end{eqnarray}

The rms radius of the intrinsic wave function of the $2s$ quartet is 
\begin{eqnarray}
&&[{\rm  rms}_{2s^4}^{\rm intr.}(R )]^2=\int d^3 r'\,r'^2\, \varrho_{2s^4}^{\rm intr.}( {\bf  r';R})
\nonumber\\&& =
3 (167333523 + 311075904\, a R^2 + 1156513536\, a^2 R^4 - 
   2022895616\, a^3 R^6
\nonumber\\&& + 4092862464\, a^4 R^8
    - 3357802496\, a^5 R^{10} + 
   1701838848\, a^6 R^{12} - 419430400\, a^7 R^{14} 
\nonumber\\&& + 50331648\, a^8 R^{16})
[8 a (24695649+14905152\, a R^2+354818304\, a^2R^4
\nonumber\\&& -876834816\, a^3R^6
+1503289344\, a^4R^8 
-1261699072\, a^5R^{10}+613416960\, a^6R^{12}
\nonumber\\&&-150994944\, a^7 R^{14}+16777216\, a^8R^{16})]^{-1}\,.
\end{eqnarray}
In the limit $R \to \infty$ the value of ${\rm rms}_{2s^4}^{\rm intr}(R )$ approaches $3/(8 a)^{1/2}$.


\begin{thebibliography}{99}


\bibitem{Matsui}
T. Matsui, Nucl. Phys. A {\bf 370}, 365 (1981).

\bibitem{RMS}
G.\ R\"opke,  L.\ M\"unchow, and H.\ Schulz,
  Nucl. Phys. A {\bf 379}, 536 (1982);\\
Phys. Lett. {\bf B 110}, 21 (1982).

\bibitem{R}
G. R\"opke, Phys. Rev. C {\bf 79}, 014002 (2009);
  Nucl. Phys. A {\bf 867}, 66 (2011);\\
 Phys. Rev. C {\bf 92}, 054001 (2015).


\bibitem{THSR} A. Tohsaki, H. Horiuchi, P. Schuck, and G. R\"opke, Phys.
Rev. Lett. {\bf 87}, 192501 (2001).

\bibitem{Bo3}
Bo Zhou, Y. Funaki, H. Horiuchi, Zhongzhou Ren, G. R\"opke, P. Schuck, A. Tohsaki,
Chang Xu, and T. Yamada,
Phys. Rev. Lett. {\bf 110}, 262501 (2013).

\bibitem{Bo}
Bo Zhou, Zhongzhou Ren, Chang Xu, Y. Funaki, T. Yamada, A.
Tohsaki, H. Horiuchi, P. Schuck, and G. R\"opke, Phys. Rev. C {\bf
86}, 014301 (2012).

\bibitem{Bo2}
Bo Zhou, Y. Funaki, H. Horiuchi, Zhongzhou Ren, G. R\"opke, P.
Schuck, A. Tohsaki, Chang Xu, and T. Yamada, Phys. Rev. C {\bf
89}, 034319 (2014).

\bibitem{Lyu}
Mengjiao Lyu, Mengjiao, Zhongzhou Ren, Bo Zhou,  {\it et al.},
 Phys. Rev. C {\bf 91}, 014313 (2015);\\
 Phys. Rev. C {\bf 93}, 054308 (2016).

\bibitem{Po}
G. R\"opke, P. Schuck, Y. Funaki, H. Horiuchi, Zhongzhou Ren, A. Tohsaki, Chang Xu, T. Yamada, and Bo Zhou,
 Phys. Rev. C {\bf 90}, 034304 (2014).

\bibitem{Xu}
Chang Xu {\it et al.}, Phys. Rev. C {\bf 93}, 011306({}R) (2016).

\bibitem{Xu1}
Chang Xu {\it et al.}, Phys. Rev. C {\bf 95}, 061306(R) (2017).

\bibitem{Mang}
H. J. Mang, Phys. Rev. {\bf 119}, 1069 (1960).

\bibitem{Arima}
I. Tonozuka and A. Arima, Nucl. Phys. A {\bf 323}, 45 (1979).

\bibitem{Delion09}
D. S. Delion, Phys. Rev. C {\bf 80}, 024310 (2009).


\bibitem{Lovas}
R. G. Lovas {\it et al.}, Phys. Rep.  {\bf 294}, 265 (1998).

\bibitem{Mirea}
M. Mirea, Eur. Phys. J. A {\bf 51}, 36 (2015).

\bibitem{Delion1}
D. S. Delion {\it et al.}, Phys. Rev. C {\bf 85}, 064306 (2012).

\bibitem{DLW}
D. S. Delion, R. J. Liotta, and R. Wyss,
Phys. Rev. C  {\bf 92}, 051301(R) (2015).

\bibitem{solidalpha}
Z. Sosin {\it et al.}, Eur. Phys. J. A {\bf 52}, 120 (2016).

\bibitem{Tarbert2014}
C. M. Tarbert {\it et al.}, Phys. Rev. Lett. {\bf 112}, 242502 (2014).

\bibitem{Qu2011}
W. W. Qu, G. L. Zhang, and X. Y. Le, Nucl. Phys. A {\bf 868} 1 (2011).

\bibitem{rms}
H. de Vries, C.W. de Jager, and C. de Vries, Atomic Data and Nuclear Data Tables {\bf 36} 495  (1987).

\bibitem{M3YReview}
G. R. Satchler and  W. G. Love, Phys. Rep. {\bf 55}, 183 (1979).



\bibitem{BBohr}
B.F. Bayman and A. Bohr, Nucl. Phys. {\bf 9},596 (1958).

\end{thebibliography}
\end{document}